\documentclass{article}
\usepackage{amssymb}
\usepackage{amsfonts}
\usepackage{amsmath}

\input{tcilatex}
\begin{document}

\title{On the relaxation to equilibrium of a quantum oscillator interacting
with a radiation field}
\author{Pierre-A. Vuillermot$^{\ast ,\ast \ast }$ \\
Center for Mathematical Studies, CEMS.UL, Faculdade de Ci\^{e}ncias,\\
Universidade de Lisboa, 1749-016 Lisboa, Portugal$^{\ast }$\\
Universit\'{e} de Lorraine, CNRS, IECL, F-54000 Nancy, France$^{\ast \ast }$%
\\
pierre.vuillermot@univ-lorraine.fr\\
ORCID Number: 0000-0003-3527-9163}
\date{}
\maketitle

\begin{abstract}
In this article we investigate from the point of view of spectral theory the
problem of relaxation to thermodynamical equilibrium of a quantum harmonic
oscillator interacting with a radiation field. Our starting point is a
system of infinitely many Pauli master equations governing the time
evolution of the occupation probabilities of the available quantum states.
The system we consider is derived from the evolution equation for the
reduced density operator obtained after the initial interaction of the
oscillator with the radiation field, the latter acting as a heat bath. We
provide a complete spectral analysis of the infinitesimal generator of the
equations, showing thereby that it generates an infinite-dimensional
dynamical system of hyperbolic type. This implies that every global solution
to the equations converges exponentially rapidly toward the corresponding
Gibbs equilibrium state. We also provide a complete spectral analysis of a
linear pencil naturally associated with the Pauli equations. All of our
considerations revolve around the notion of compactness, more specifically
around the existence of a new compact embedding result involving a space of
sequences of Sobolev type into a weighted space of square-integrable
summable sequences.



\ \ \ \ \ \textbf{Abbreviated Title: Relaxation to Equilibrium}
\end{abstract}

\section{Introduction and outline}

In this article we investigate the relaxation to thermodynamical equilibrium
of a one-dimensional quantum harmonic oscillator interacting with a
radiation field, for instance a quantized electromagnetic field in a vacuum
or lattice vibrations in a crystal, acting as a heat bath. More
specifically, we are primarily interested in the long time behavior of the
occupation probabilities associated with each quantum energy state of the
oscillator after its initial interaction with the bath. The system governing
the evolution of those probabilities consists of infinitely many master
equations, that is, one such equation for each quantum state. The equations
we consider are typical gain-loss equations whose structure is already
apparent in \cite{pauli}, \textit{albeit} in a very different context. They
represent a Markovian approximation to a more general evolution system that
takes the form of an integro-differential equation with a memory term for
the reduced density operator under consideration. Accordingly, we organize
our article in the following way: In Section 2 we define the class of master
equations to be analyzed and prove the self-adjointness of their generator
on a dense domain of a suitable Hilbert space. We proceed by proving that
the generator in question has a compact resolvent and thereby a discrete
spectrum, which requires a concrete realization of its domain as a weighted
Sobolev space of sequences compactly embedded in the given Hilbert space.
This allows us to prove that the system generates a dynamical system of
hyperbolic type, and from there we obtain the spectral resolution of the
corresponding evolution semi-group, which is holomorphic and contracting.
Knowing this, we then prove that any initial probability distribution
remains a probability distribution for all times, which eventually converges
exponentially rapidly, as time becomes large, to the Gibbs equilibrium
distribution characterized by the temperature of the bath. In Section 3 we
determine the spectrum of a linear pencil naturally associated with the
system of equations analyzed in Section 2. We conclude the article with two
appendices. In the first one we prove a generalization of the compactness
result used in Sections 2 and 3 along with some consequences. In the second
one we briefly explain how the interaction of the harmonic oscillator with
the radiation field leads to the considered system of master equations,
starting out with the evolution equation for the reduced density operator of
the harmonic oscillator after its initial interaction with the heat bath and
the elimination of the bath variables.

Throughout this work we make a point of giving detailed proofs of all
statements, with appropriate references whenever necessary.

Needless to say, much is known regarding the properties of a quantum
harmonic oscillator in a heat bath and their physical consequences (see,
e.g. \cite{daviesbis}, \cite{haake} or \cite{vankampen} and some of the
references therein). Thus, the focus of our approach to the problem of
relaxation by means of spectral arguments is not so much about the results
they imply as it is about the strategy to obtain them. Indeed, to the best
of our knowledge, the spectral analysis of the generator of the master
equations viewed as an unbounded self-adjoint operator on some suitable
weighted sequence space and based on compactness arguments is new, as is its
relation to the spectral theory of linear pencils. As such, our method may
then be considered as complementary to the existing ones, particularly to
those set forth in \cite{daviesbis} and \cite{daviester}, which make use of
completely different procedures, and thus renders our contribution
self-contained.

For other applications of master equations in non-equilibrium statistical
mechanics, we refer the reader for instance to [$3,11,16,22$] and their
references.

\section{A spectral theorem for a class of infinitely many Pauli master
equations}

Our starting point to implement the above program is the system of Pauli
master equations given by%
\begin{eqnarray}
\frac{dp_{\mathsf{m}}(t)}{dt} &=&\dsum\limits_{\mathsf{n=0}}^{+\infty
}\left( r_{\mathsf{m,n}}p_{\mathsf{n}}(t)-r_{\mathsf{n,m}}p_{\mathsf{m}%
}(t)\right) ,\text{ \ }t\in \left( 0,+\infty \right)  \notag \\
p_{\mathsf{m}}(0) &=&p_{\mathsf{m}}^{\ast }  \label{masterequations}
\end{eqnarray}%
for every $\mathsf{m}\in \mathbb{N}$, where $(p_{\mathsf{m}}^{\ast })$
stands for any sequence of initial conditions satisfying%
\begin{equation}
p_{\mathsf{m}}^{\ast }\geq 0,\text{ \ }\sum_{\mathsf{m}=0}^{+\infty }p_{%
\mathsf{m}}^{\ast }=1,  \label{probabilities}
\end{equation}%
and where the transition rates from level $\mathsf{n}$ to level $\mathsf{m}$
are%
\begin{equation}
r_{\mathsf{m,n}}=\left\{ 
\begin{array}{c}
-(\rho \mathsf{m}+\sigma (\mathsf{m}+1))\text{ \ \ for \ }\mathsf{m=n,} \\ 
\\ 
\rho \mathsf{n}\delta _{\mathsf{m,n-1}}+\sigma (\mathsf{n}+1)\delta _{%
\mathsf{m,n+1}}\text{ \ \ for }\mathsf{m\neq n}%
\end{array}%
\right.  \label{rates}
\end{equation}%
with $\rho ,\sigma >0$. It will soon be reminded that the functions $%
t\mapsto $ $p_{\mathsf{m}}(t)$ are the occupation probabilities of level $%
\mathsf{m}$ at time $t$ of the harmonic oscillator after its initial
interaction with the heat bath.

While system (\ref{masterequations}) takes the form%
\begin{eqnarray}
\frac{dp_{\mathsf{m}}(t)}{dt} &=&\rho (\mathsf{m}+1)p_{\mathsf{m+1}%
}(t)+\sigma \mathsf{m}p_{\mathsf{m-1}}(t)-\left( \rho \mathsf{m}+\sigma
\left( \mathsf{m}+1\right) \right) p_{\mathsf{m}}(t),  \notag \\
p_{\mathsf{m}}(0) &=&p_{\mathsf{m}}^{\ast },  \label{masterequationsbis}
\end{eqnarray}%
as a consequence of (\ref{rates}), we note that the right-hand side of the
preceding equation is ill-defined for $\mathsf{m=0}$ because of the presence
of $p_{\mathsf{m-1}}(t)$. Strictly speaking, it would indeed be more
accurate to require (\ref{masterequationsbis}) for $\mathsf{m}\geqslant 1$
while imposing%
\begin{eqnarray*}
\frac{dp_{\mathsf{0}}(t)}{dt} &=&\rho p_{\mathsf{1}}(t)-\sigma p_{\mathsf{0}%
}(t), \\
p_{\mathsf{0}}(0) &=&p_{\mathsf{0}}^{\ast }
\end{eqnarray*}%
for $\mathsf{m=0}$. However, since the coefficient of $p_{\mathsf{m-1}}(t)$
vanishes when $\mathsf{m=0}$ we will keep using (\ref{masterequationsbis})
for the sake of convenience and simplicity of the notation. A similar remark
applies to the second line of (\ref{rates}).

\bigskip

Before we make a first contact with statistical mechanics, we have to
determine the unique stationary solution to (\ref{masterequationsbis}):

\bigskip

\textbf{Lemma 1. }\textit{Aside from the positivity of }$\sigma $ \textit{and%
} $\rho $, \textit{let us assume that }$\sigma <\rho $. \textit{Then} 
\textit{system (\ref{masterequationsbis}) possesses exactly one stationary
solution} $\mathsf{p}_{\mathsf{S}}$ \textit{given by}%
\begin{equation}
p_{\mathsf{S},\mathsf{m}}=\left( \frac{\sigma }{\rho }\right) ^{\mathsf{m}%
}\left( 1-\frac{\sigma }{\rho }\right)  \label{stationary}
\end{equation}%
\textit{for every} $\mathsf{m}\in \mathbb{N}$.

\bigskip

\textbf{Proof. }Setting the right-hand side of (\ref{masterequationsbis})
equal to zero we get the recurrence relation%
\begin{equation*}
p_{\mathsf{S},\mathsf{m+1}}=\frac{\left( \rho \mathsf{m}+\sigma \left( 
\mathsf{m}+1\right) \right) p_{\mathsf{S},\mathsf{m}}-\sigma \mathsf{m}p_{%
\mathsf{S},\mathsf{m-1}}}{\rho (\mathsf{m}+1)},
\end{equation*}%
so that we obtain%
\begin{equation*}
p_{\mathsf{S},\mathsf{m}}=\left( \frac{\sigma }{\rho }\right) ^{\mathsf{m}%
}p_{\mathsf{S},\mathsf{0}}
\end{equation*}%
for every $\mathsf{m}\in \mathbb{N}$ from an easy induction argument. Since $%
\sigma <\rho $, the result then follows from the normalization condition (%
\ref{probabilities}). \ \ $\blacksquare $

\bigskip

From now on we identify the energy spectrum of the one-dimensional quantum
harmonic oscillator with $\lambda _{\mathsf{m}}=\mathsf{m}\in \mathbb{N}$,
which only requires one to rescale and shift the corresponding Hamiltonian
operator in an irrelevant way. When the oscillator is in equilibrium with
the heat bath at inverse temperature $\beta =\left( k_{\mathsf{B}}T\right)
^{-1}>0$, where $k_{\mathsf{B}}$ stands for Boltzmann's constant, the
corresponding Gibbs probability distribution $\mathsf{p}_{\beta ,\mathsf{%
Gibbs}}$ is then%
\begin{equation}
p_{\beta ,\mathsf{Gibbs},\mathsf{m}}=Z_{\beta }^{-1}\exp \left[ -\beta 
\mathsf{m}\right]  \label{gibbs}
\end{equation}%
where 
\begin{equation}
Z_{\beta }=\sum_{\mathsf{m=0}}^{+\infty }\exp \left[ -\beta \mathsf{m}\right]
=\left( 1-\exp \left[ -\beta \right] \right) ^{-1}
\label{partittionfunction}
\end{equation}%
is the associated partition function. We can easily arrange for $\mathsf{p}%
_{\beta ,\mathsf{Gibbs}}$ to be a trivial solution to (\ref%
{masterequationsbis}) when the initial condition is chosen as $p_{\mathsf{m}%
}^{\ast }=p_{\beta ,\mathsf{Gibbs},\mathsf{m}}$ for every $\mathsf{m}\in 
\mathbb{N}$. Indeed the following result holds:

\bigskip

\textbf{Lemma 2. }\textit{The rates (\ref{rates}) satisfy the so-called
detailed balance conditions}%
\begin{equation}
r_{\mathsf{m,n}}p_{\beta ,\mathsf{Gibbs},\mathsf{n}}=r_{\mathsf{n,m}%
}p_{\beta ,\mathsf{Gibbs},\mathsf{m}}  \label{balance condition}
\end{equation}%
\textit{with respect to (\ref{gibbs}) for all }$\mathsf{m,n}\in \mathbb{N}$ 
\textit{if, and only if,}%
\begin{equation}
\frac{\sigma }{\rho }=\exp \left[ -\beta \right] .  \label{relation}
\end{equation}%
\textit{In this case we have}%
\begin{equation}
\mathsf{p}_{\mathsf{S}}=\mathsf{p}_{\beta ,\mathsf{Gibbs}}.
\label{equilibrium}
\end{equation}

\bigskip

\textbf{Proof. }It is sufficient to assume $\mathsf{m\neq n}$, in which case
(\ref{balance condition}) reads%
\begin{eqnarray*}
&&\left( \rho \mathsf{n}\delta _{\mathsf{m,n-1}}+\sigma (\mathsf{n}+1)\delta
_{\mathsf{m,n+1}}\right) \exp \left[ -\beta \mathsf{n}\right] \\
&=&\left( \rho \mathsf{m}\delta _{\mathsf{n,m-1}}+\sigma (\mathsf{m}%
+1)\delta _{\mathsf{n,m+1}}\right) \exp \left[ -\beta \mathsf{m}\right] ,
\end{eqnarray*}%
from which (\ref{relation}) follows when $\mathsf{m=n-1}$ or when $\mathsf{%
m=n+1}$. The converse is equally trivial and the substitution of (\ref%
{relation}) into (\ref{stationary}) then gives (\ref{equilibrium}). \ \ $%
\blacksquare $

\bigskip

The detailed balance conditions (\ref{balance condition}) thus imply that
the parameters $\sigma $ and $\rho $ are not independent, and as a
consequence the unique stationary solution to (\ref{masterequationsbis}) is
an equilibrium solution in the thermodynamical sense. Before proceeding, we
note that (\ref{rates}) becomes%
\begin{equation}
r_{\mathsf{m,n}}=\left\{ 
\begin{array}{c}
-\rho (\mathsf{m}+\exp \left[ -\beta \right] (\mathsf{m}+1))\text{ \ \ for \ 
}\mathsf{m=n,} \\ 
\\ 
\rho \left( \mathsf{n}\delta _{\mathsf{m,n-1}}+\exp \left[ -\beta \right] (%
\mathsf{n}+1)\delta _{\mathsf{m,n+1}}\text{ }\right) \text{ \ \ for }\mathsf{%
m\neq n}%
\end{array}%
\right.  \label{condition}
\end{equation}%
for every $\mathsf{m}$ as a consequence of (\ref{relation}), so that we may
rewrite the equation in (\ref{masterequationsbis}) as%
\begin{eqnarray}
&&\frac{dp_{\mathsf{m}}(t)}{dt}  \label{masterequationster} \\
&=&\rho \left\{ (\mathsf{m}+1)p_{\mathsf{m+1}}(t)+\exp \left[ -\beta \right] 
\mathsf{m}p_{\mathsf{m-1}}(t)-\left( \mathsf{m}+\exp \left[ -\beta \right]
\left( \mathsf{m}+1\right) \right) p_{\mathsf{m}}(t)\right\} .  \notag
\end{eqnarray}

Regarding the question of how the dynamics of the coupled system consisting
of the harmonic oscillator and the radiation field acting as a heat bath
leads to (\ref{masterequationster}), we refer the reader for instance to [$%
12,17,26$-$28$], and to Appendix B for a brief summary. In a nutshell, as in 
\cite{haake} and in Sections 3 and 4 of Chapter XVII in \cite{vankampen},
the bath is modelled by an infinite system of quantum harmonic oscillators
whose frequencies satisfy certain conditions. The time evolution of the
density operator describing that coupling is then governed by the laws of
Quantum Mechanics. Subsequently, various methods are used to eliminate the
bath variables to eventually obtain the evolution equation of the so-called
reduced density operator, that is, the density operator of the given
oscillator after its initial interaction with the bath. Among others, one of
those methods is the celebrated Projection Technique devised in [$17,27,28$%
]. As a consequence and within a Markovian approximation, equation (\ref%
{masterequationster}) emerges naturally with the functions $t\rightarrow $ $%
p_{\mathsf{m}}(t)$ being defined as the diagonal matrix elements of the
reduced density operator with respect to the $\mathsf{m}^{\mathsf{th}}$
eigenfunction of the oscillator. We refer the reader for instance to [$5$-$8$%
] for a rigorous implementation of some of those formal developments.

We devote the remaining part of this section to proving that $\mathsf{p}%
_{\beta ,\mathsf{Gibbs}}$ is in fact the global exponential attractor to (%
\ref{masterequationster}), more specifically that the dynamical system
generated by (\ref{masterequationster}) is hyperbolic, which implies that
its solution converges exponentially rapidly to $\mathsf{p}_{\beta ,\mathsf{%
Gibbs}}$ as $\tau \rightarrow +\infty $ in a suitable topology. We first
introduce a suitable functional space in order to make a self-adjoint
realization out of the formal expression on the right-hand side of (\ref%
{masterequationster}). Let $w_{\beta }:=(w_{\beta ,\mathsf{m}})$ be the
sequence of weights given by $w_{\beta ,\mathsf{m}}=\exp \left[ \beta 
\mathsf{m}\right] $ for each $\mathsf{m}\in \mathbb{N}$, and let us consider
the separable Hilbert space $l_{\mathbb{C},w_{\beta }}^{2}$ consisting of
all complex sequences $\mathsf{p=(}p_{\mathsf{m}}\mathsf{)}$ satisfying%
\begin{equation}
\left\Vert \mathsf{p}\right\Vert _{2,w_{\beta }}^{2}:=\sum_{\mathsf{m}%
=0}^{+\infty }w_{\beta ,\mathsf{m}}\left\vert p_{\mathsf{m}}\right\vert
^{2}<+\infty ,  \label{norm}
\end{equation}%
endowed with the usual algebraic operations and the sesquilinear form%
\begin{equation}
\left( \mathsf{p,q}\right) _{2,w_{\beta }}:=\sum_{\mathsf{m}=0}^{+\infty
}w_{\beta ,\mathsf{m}}p_{\mathsf{m}}\bar{q}_{\mathsf{m}}.
\label{innerproduct}
\end{equation}%
Let us consider the sequence $\left( \mathsf{f}_{\mathsf{m}}\right) $ of
elements in$\mathsf{\ }l_{\mathbb{C},w_{\beta }}^{2}$ defined by $\mathsf{f}%
_{\mathsf{m}}=w_{\beta ,\mathsf{m}}^{-\frac{1}{2}}\mathsf{e}_{\mathsf{m}}$
where $\left( \mathsf{e}_{\mathsf{m}}\right) _{\mathsf{n}}=\delta _{\mathsf{m%
},\mathsf{n}}$ for all $\mathsf{m,n}\in \mathbb{N}$. It is easily seen that $%
\left( \mathsf{f}_{\mathsf{m}}\right) $ provides an orthonormal basis of
that space. Moreover we have%
\begin{equation}
\sum_{\mathsf{m=0}}^{+\infty }r_{\mathsf{m,n}}=0  \label{conditionbis}
\end{equation}%
for every $\mathsf{n}\in \mathbb{N}$. Then the following result holds:

\bigskip

\textbf{Proposition 1.}\textit{\ Let us define the mapping }$R$\textit{\ by}%
\begin{equation}
\left( R\mathsf{p}\right) _{\mathsf{m}}=\sum_{\mathsf{n=0}}^{+\infty }r_{%
\mathsf{m},\mathsf{n}}p_{\mathsf{n}}.  \label{mapping}
\end{equation}%
\textit{Then }$R$\textit{\ is an unbounded symmetric operator on the dense
domain }$D(R)\subset l_{\mathbb{C},w_{\beta }}^{2}$\textit{\ consisting of
all finite linear combinations of the}\textsf{\ }$\mathsf{f}_{\mathsf{m}}$%
\textit{.}

\bigskip

\textbf{Proof.} By linearity it is sufficient to prove that%
\begin{equation*}
\left( R\mathsf{f}_{\mathsf{m}},\mathsf{f}_{\mathsf{n}}\right) _{2,w_{\beta
}}=\left( \mathsf{f}_{\mathsf{m}},R\mathsf{f}_{\mathsf{n}}\right)
_{2,w_{\beta }}
\end{equation*}%
for all $\mathsf{m,n}\in \mathbb{N}$. Noting that $\left( \mathsf{f}_{%
\mathsf{m}}\right) _{\mathsf{k}}=w_{\beta ,\mathsf{m}}^{-\frac{1}{2}}\delta
_{\mathsf{m},\mathsf{k}}$ and $\left( R\mathsf{f}_{\mathsf{n}}\right) _{%
\mathsf{k}}=w_{\beta ,\mathsf{n}}^{-\frac{1}{2}}r_{\mathsf{k,n}}$ we have%
\begin{equation*}
\left( R\mathsf{f}_{\mathsf{m}},\mathsf{f}_{\mathsf{n}}\right) _{2,w_{\beta
}}=w_{\beta ,\mathsf{m}}^{-\frac{1}{2}}w_{\beta ,\mathsf{n}}^{\frac{1}{2}}r_{%
\mathsf{n,m}}
\end{equation*}%
and%
\begin{equation*}
\left( \mathsf{f}_{\mathsf{m}},R\mathsf{f}_{\mathsf{n}}\right) _{2,w_{\beta
}}=w_{\beta ,\mathsf{m}}^{\frac{1}{2}}w_{\beta ,\mathsf{n}}^{-\frac{1}{2}}r_{%
\mathsf{m,n}}
\end{equation*}%
respectively, so that both expressions are equal by virtue of the detailed
balance conditions. Furthermore we have%
\begin{equation*}
\sum_{\mathsf{k}=0}^{+\infty }w_{\beta ,\mathsf{k}}\left\vert \left( R%
\mathsf{f}_{\mathsf{m}}\right) _{\mathsf{k}}\right\vert ^{2}=w_{\beta ,%
\mathsf{m}}^{-1}\sum_{\mathsf{k}=0}^{+\infty }w_{\beta ,\mathsf{k}}r_{%
\mathsf{k,m}}^{2},
\end{equation*}%
which, by taking (\ref{condition}) and the definition of the weights into
account, leads to%
\begin{eqnarray*}
\left\Vert R\mathsf{f}_{\mathsf{m}}\right\Vert _{2,w_{\beta }}^{2} &=&\rho
^{2}\left\{ (\mathsf{m}+\exp \left[ -\beta \right] (\mathsf{m}+1))^{2}+\exp %
\left[ -\beta \right] \left( 2\mathsf{m}^{2}+2\mathsf{m}+1\right) \right\} \\
&\geq &\rho ^{2}\exp \left[ -\beta \right] \mathsf{m}^{2}\rightarrow +\infty
\end{eqnarray*}%
as $\mathsf{m}\rightarrow +\infty $, so that $R$ is indeed unbounded. \ \ $%
\blacksquare $

\bigskip

\textsc{Remark.} It is essential to have a functional space endowed with an
appropriately weighted norm in order to make a symmetric operator out of (%
\ref{mapping}) by means of the detailed balance conditions (\ref{balance
condition}) (see, e.g., Section 7 in Chapter V of \cite{vankampen} for an
informal discussion of this idea in a general context, which has been around
for some time). Had we chosen the usual unweighted space $l_{\mathbb{C}}^{2}$
consisting of all square-integrable complex sequences instead, the operator $%
R$ would not have been symmetric, the spectral analysis of its possible non
self-adjoint extensions being thereby definitely more complicated. Such a
situation was thoroughly investigated in \cite{boeglivuillermot}, \textit{%
albeit} with a different operator whose choice was motivated by the works on
stochastic thermodynamics of chemical reaction systems set forth in \cite%
{tomeoliveira} and \cite{tomeoliveirabis}.

\bigskip

It is straightforward to check that $\mathsf{p}_{\beta ,\mathsf{Gibbs}}\in
l_{\mathbb{C},w_{\beta }}^{2}$, but unfortunately $\mathsf{p}_{\beta ,%
\mathsf{Gibbs}}\notin D(R)$. In fact, the Fourier coefficients of $\mathsf{p}%
_{\beta ,\mathsf{Gibbs}}$ along the basis $\left( \mathsf{f}_{\mathsf{m}%
}\right) $ are%
\begin{equation*}
\left( \mathsf{p}_{\beta ,\mathsf{Gibbs}},\mathsf{f}_{\mathsf{m}}\right)
_{2,w_{\beta }}=Z_{\beta }^{-1}w_{\beta ,\mathsf{m}}^{-\frac{1}{2}}>0
\end{equation*}%
for every $\mathsf{m}\in \mathbb{N}$, so that%
\begin{equation*}
\mathsf{p}_{\beta ,\mathsf{Gibbs}}=Z_{\beta }^{-1}\sum_{\mathsf{m}%
=0}^{+\infty }w_{\beta ,\mathsf{m}}^{-\frac{1}{2}}\mathsf{f}_{\mathsf{m}}
\end{equation*}%
is not a finite linear combination of the $\mathsf{f}_{\mathsf{m}}$.
However, let $R^{\ast }$ be the adjoint of $R$. As is well known, the
operator $R^{\ast }$ provides a closed extension of $R$, namely, $R\subseteq
R^{\ast }$, so that its domain $D(R^{\ast })$ is dense in $l_{\mathbb{C}%
,w_{\beta }}^{2}$ as well. Moreover, recall that $\mathsf{q}\in D(R^{\ast })$
means there exists a unique $\mathsf{q}^{\ast }\in l_{\mathbb{C},w_{\beta
}}^{2}$ such that%
\begin{equation*}
\left( R\mathsf{p},\mathsf{q}\right) _{2,w_{\beta }}=\left( \mathsf{p},%
\mathsf{q}^{\ast }\right) _{2,w_{\beta }}
\end{equation*}%
for every $\mathsf{p}\in D(R)$, with $\mathsf{q}^{\ast }=R^{\ast }\mathsf{q}$
(for all these notions and various relations among them see, e.g., \cite%
{conway}). Then the following statement holds:

\bigskip

\textbf{Lemma 3. }\textit{We have} $\mathsf{p}_{\beta ,\mathsf{Gibbs}}\in
D(R^{\ast })$, \textit{hence }$R^{\ast }$\textit{\ is a strict extension of }%
$R$\textit{. Moreover}%
\begin{equation}
R^{\ast }\mathsf{p}_{\beta ,\mathsf{Gibbs}}=0,  \label{eigenequation}
\end{equation}%
\textit{so that }$\nu =0$\textit{\ is an eigenvalue of }$R^{\ast }$\textit{.}

\bigskip

\textbf{Proof. }We have%
\begin{equation*}
\left( R\mathsf{f}_{\mathsf{m}},\mathsf{p}_{\beta ,\mathsf{Gibbs}}\right)
_{2,w_{\beta }}=Z_{\beta }^{-1}w_{\beta ,\mathsf{m}}^{-\frac{1}{2}}\sum_{%
\mathsf{k}=0}^{+\infty }r_{\mathsf{k,m}}=0
\end{equation*}%
for every $\mathsf{m}\in \mathbb{N}$ by virtue of (\ref{conditionbis}), so
that by linearity 
\begin{equation*}
\left( R\mathsf{p},\mathsf{p}_{\beta ,\mathsf{Gibbs}}\right) _{2,w_{\beta
}}=0
\end{equation*}%
for every $\mathsf{p}\in D(R)$, which may be written as%
\begin{equation*}
\left( R\mathsf{p},\mathsf{p}_{\beta ,\mathsf{Gibbs}}\right) _{2,w_{\beta
}}=\left( \mathsf{p},\mathsf{p}_{\beta ,\mathsf{Gibbs}}^{\ast }\right)
_{2,w_{\beta }}
\end{equation*}%
with $\mathsf{p}_{\beta ,\mathsf{Gibbs}}^{\ast }=0$. Hence (\ref%
{eigenequation}) holds and the first part of the statement follows from the
fact that $\mathsf{p}_{\beta ,\mathsf{Gibbs}}\notin D(R)$. \ \ $\blacksquare 
$

\bigskip

Let us now determine the action of $R^{\ast }$ on elements of $D(R^{\ast })$%
. Not surprisingly we obtain:

\bigskip

\textbf{Lemma 4. }\textit{For every }$\mathsf{q}\in D(R^{\ast })$\textit{\
we have}%
\begin{equation}
\left( R^{\ast }\mathsf{q}\right) _{\mathsf{m}}=\rho \left\{ \left( \mathsf{m%
}+1\right) q_{\mathsf{m+1}}+\exp \left[ -\beta \right] \mathsf{m}q_{\mathsf{%
m-1}}-(\mathsf{m}+\exp \left[ -\beta \right] (\mathsf{m}+1))q_{\mathsf{m}%
}\right\}  \label{adjointpremier}
\end{equation}%
\textit{for every} $\mathsf{m}\in \mathbb{N}$, \textit{with} 
\begin{equation}
\sum_{\mathsf{m}=0}^{+\infty }w_{\beta ,\mathsf{m}}\left\vert \left( R^{\ast
}\mathsf{q}\right) _{\mathsf{m}}\right\vert ^{2}<+\infty .
\label{squareintegrable}
\end{equation}

\bigskip

\textbf{Proof. }In order to determine $\mathsf{q}^{\ast }=R^{\ast }\mathsf{q}
$ we must have in particular%
\begin{equation}
\left( R\mathsf{f}_{\mathsf{m}},\mathsf{q}\right) _{2,w_{\beta }}=\left( 
\mathsf{f}_{\mathsf{m}},\mathsf{q}^{\ast }\right) _{2,w_{\beta }}
\label{adjoint}
\end{equation}%
for every $\mathsf{m}\in \mathbb{N}$. But using (\ref{condition}) and the
definition of the weights once again we get%
\begin{eqnarray}
&&\left( R\mathsf{f}_{\mathsf{m}},\mathsf{q}\right) _{2,w_{\beta }}
\label{innerproductbis} \\
&=&\rho \exp \left[ \frac{\beta \mathsf{m}}{2}\right] \left\{ \left( \mathsf{%
m}+1\right) \bar{q}_{\mathsf{m+1}}+\exp \left[ -\beta \right] \mathsf{m}\bar{%
q}_{\mathsf{m-1}}-(\mathsf{m}+\exp \left[ -\beta \right] (\mathsf{m}+1))\bar{%
q}_{\mathsf{m}}\right\} ,  \notag
\end{eqnarray}%
while on the other hand we have%
\begin{equation*}
\left( \mathsf{f}_{\mathsf{m}},\mathsf{q}^{\ast }\right) _{2,w_{\beta
}}=\exp \left[ \frac{\beta \mathsf{m}}{2}\right] \bar{q}_{\mathsf{m}}^{\ast
}.
\end{equation*}%
Therefore, solving (\ref{adjoint}) for $q_{\mathsf{m}}^{\ast }$ gives the
desired result while (\ref{squareintegrable}) expresses the requirement $%
R^{\ast }\mathsf{q}\in l_{\mathbb{C},w_{\beta }}^{2}$. \ \ $\blacksquare $

\bigskip

The operator $R^{\ast }$ turns out to be the desired self-adjoint
realization we need to deal with the right-hand side of (\ref%
{masterequationster}):

\bigskip

\textbf{Proposition 2.} \textit{The unbounded operator }$R^{\ast }$\textit{\
is self-adjoint on }$D(R^{\ast })$\textit{. In other words, the operator }$R$%
\textit{\ is essentially self-adjoint on }$D(R)$\textit{.}

\bigskip

\textbf{Proof. }Since $R^{\ast \ast }:=\left( R^{\ast }\right) ^{\ast }$ is
the smallest closed extension of $R$ we have $R\subseteq R^{\ast \ast
}\subseteq R^{\ast }$, so that it is sufficient to prove the symmetry of $%
R^{\ast }$ on\textit{\ }$D(R^{\ast })$ for then%
\begin{equation*}
R\subseteq R^{\ast \ast }\subseteq R^{\ast }\subseteq R^{\ast \ast },
\end{equation*}%
which implies $R\subset R^{\ast }=R^{\ast \ast }$, the strict inclusion
following from the information we already have. But from (\ref%
{adjointpremier}) we get%
\begin{eqnarray}
&&\left( R^{\ast }\mathsf{p},\mathsf{q}\right) _{2,w_{\beta }}  \notag \\
&=&\rho \dsum\limits_{\mathsf{m=0}}^{+\infty }\left( \exp \left[ \beta 
\mathsf{m}\right] (\mathsf{m+1)}p_{\mathsf{m+1}}\bar{q}_{\mathsf{m}}+\exp %
\left[ \beta (\mathsf{m-1})\right] \mathsf{m}p_{\mathsf{m-1}}\bar{q}_{%
\mathsf{m}}\right)  \label{symmetry} \\
&&-\rho \dsum\limits_{\mathsf{m=0}}^{+\infty }\exp \left[ \beta \mathsf{m}%
\right] \left( \mathsf{m+}\exp \left[ -\beta \right] \left( \mathsf{m+1}%
\right) \right) p_{\mathsf{m}}\bar{q}_{\mathsf{m}}  \notag
\end{eqnarray}%
for all $\mathsf{p,q}\in D(R^{\ast }),$ while%
\begin{eqnarray}
&&\left( \mathsf{p},R^{\ast }\mathsf{q}\right) _{2,w_{\beta }}  \notag \\
&=&\rho \dsum\limits_{\mathsf{m=0}}^{+\infty }\left( \exp \left[ \beta 
\mathsf{m}\right] (\mathsf{m+1)}p_{\mathsf{m}}\bar{q}_{\mathsf{m+1}}+\exp %
\left[ \beta (\mathsf{m-1})\right] \mathsf{m}p_{\mathsf{m}}\bar{q}_{\mathsf{%
m-1}}\right)  \label{symmetrybis} \\
&&-\rho \dsum\limits_{\mathsf{m=0}}^{+\infty }\exp \left[ \beta \mathsf{m}%
\right] \left( \mathsf{m+}\exp \left[ -\beta \right] \left( \mathsf{m+1}%
\right) \right) p_{\mathsf{m}}\bar{q}_{\mathsf{m}},  \notag
\end{eqnarray}%
from which we easily see that 
\begin{equation*}
\left( R^{\ast }\mathsf{p},\mathsf{q}\right) _{2,w_{\beta }}=\left( \mathsf{p%
},R^{\ast }\mathsf{q}\right) _{2,w_{\beta }}
\end{equation*}%
by means of suitable shifts in the summation indices wherever necessary. \ \ 
$\blacksquare $

\bigskip

Next, we prove a series of results which will eventually lead to the
determination of the spectrum of $R^{\ast }$. We begin with the following
statement, which makes the information already provided by (\ref%
{eigenequation}) more precise:

\bigskip

\textbf{Proposition 3.} \textit{Let} $\mathsf{q}\in D(R^{\ast })$\textit{\
be such that }$R^{\ast }\mathsf{q}=0$\textit{. Then we have}%
\begin{equation}
\mathsf{q=}q_{0}Z_{\beta }\mathsf{p}_{\beta ,\mathsf{Gibbs}}  \label{kernel}
\end{equation}%
\textit{for some} $q_{0}\in \mathbb{C}$, \textit{where }$Z_{\beta }$\textit{%
\ is the partition function defined in (\ref{gibbs}). Thus, }$\nu =0$\textit{%
\ is a simple eigenvalue of }$R^{\ast }$ \textit{whose corresponding
eigenspace is generated by }$\mathsf{p}_{\beta ,\mathsf{Gibbs}}$.

\bigskip

\textbf{Proof.} Requiring $\left( R^{\ast }\mathsf{q}\right) _{\mathsf{m}}=0$
for every $\mathsf{m}$ from (\ref{adjointpremier}) gives%
\begin{equation}
q_{\mathsf{m}+1}=\frac{\left( \mathsf{m}+\exp \left[ -\beta \right] (\mathsf{%
m}+1)\right) q_{\mathsf{m}}-\exp \left[ -\beta \right] \mathsf{m}q_{\mathsf{m%
}-1}}{\mathsf{m+1}} ,  \label{recurrence}
\end{equation}%
hence $q_{1}=\exp \left[ -\beta \right] q_{0}$ for some $q_{0}\in \mathbb{C}$
when $\mathsf{m=}$ $0$. Therefore, if we assume $q_{\mathsf{m}}=\exp \left[
-\beta \mathsf{m}\right] q_{0}$ for some $\mathsf{m}$ then (\ref{recurrence}%
) implies 
\begin{eqnarray*}
q_{\mathsf{m}+1} &=&\frac{\left( \mathsf{m}+\exp \left[ -\beta \right] (%
\mathsf{m}+1)\right) \exp \left[ -\beta \mathsf{m}\right] q_{0}-\mathsf{m}%
\exp \left[ -\beta \mathsf{m}\right] q_{0}}{\mathsf{m+1}} \\
&=&\exp \left[ -\beta \left( \mathsf{m+1}\right) \right] q_{0},
\end{eqnarray*}%
so that%
\begin{equation*}
q_{\mathsf{m}}=\exp \left[ -\beta \mathsf{m}\right] q_{0}
\end{equation*}%
holds for every $\mathsf{m}\in \mathbb{N}$, which gives (\ref{kernel})
according to (\ref{gibbs}). \ \ $\blacksquare $\ \ 

\bigskip

Since $R^{\ast }$ is self-adjoint, its entire spectrum is a subset of the
real axis. The following statement will provide additional information about
its localization:

\bigskip

\textbf{Lemma 5.} \textit{The operator }$R^{\ast }$ \textit{is negative
semi-definite. More specifically, for each} $\mathsf{q}\in D(R^{\ast })$ 
\textit{we have}%
\begin{equation}
\left( R^{\ast }\mathsf{q},\mathsf{q}\right) _{2,w_{\beta }}<0
\label{quadraticform}
\end{equation}%
\textit{unless }$\mathsf{q}$ \textit{is of the form (\ref{kernel}). In
particular, the strict inequality (\ref{quadraticform}) holds for every
non-zero }$\mathsf{q}\in D(R^{\ast })$ \textit{in the orthogonal complement
of the one-dimensional subspace generated by }$\mathsf{p}_{\beta ,\mathsf{%
Gibbs}}$. \textit{Moreover, if }$\nu \in \mathbb{C}$ \textit{with }$\func{Re}%
\nu >0$ \textit{then}%
\begin{equation}
\left\Vert \left( R^{\ast }-\nu \right) \mathsf{q}\right\Vert _{2,w_{\beta
}}\geq \left\vert \nu \right\vert \left\Vert \mathsf{q}\right\Vert
_{2,w_{\beta }}.  \label{estimate}
\end{equation}

\bigskip

\textbf{Proof.} Since both sums on the right-hand side of the equality in (%
\ref{symmetry}) are real when $\mathsf{p}=\mathsf{q}$, the proof of (\ref%
{quadraticform}) amounts to proving that%
\begin{eqnarray}
&&\dsum\limits_{\mathsf{m=0}}^{+\infty }\left( \exp \left[ \beta \mathsf{m}%
\right] (\mathsf{m+1)}q_{\mathsf{m+1}}\bar{q}_{\mathsf{m}}+\exp \left[ \beta
\left( \mathsf{m-1}\right) \right] \mathsf{m}q_{\mathsf{m-1}}\bar{q}_{%
\mathsf{m}}\right)  \notag \\
&<&\dsum\limits_{\mathsf{m=0}}^{+\infty }\exp \left[ \beta \mathsf{m}\right]
\left( \mathsf{m+}\exp \left[ -\beta \right] \left( \mathsf{m+1}\right)
\right) \left\vert q_{\mathsf{m}}\right\vert ^{2}  \label{estimatebis}
\end{eqnarray}%
for every $\mathsf{q}\in D(R^{\ast })$, unless\textit{\ }$\mathsf{q}$ is of
the form (\ref{kernel})\textit{. }In order to achieve that we proceed by
using Cauchy's inequality with epsilon to deal with the left-hand side of
the preceding inequality. We begin by writing%
\begin{eqnarray*}
&&\dsum\limits_{\mathsf{m=0}}^{+\infty }\exp \left[ \beta \mathsf{m}\right] (%
\mathsf{m+1)}\left\vert q_{\mathsf{m+1}}\bar{q}_{\mathsf{m}}\right\vert \\
&=&\dsum\limits_{\mathsf{m=0}}^{+\infty }\exp \left[ \frac{\beta }{2}\mathsf{%
m}\right] (\mathsf{m+1)}^{\frac{1}{2}}\left\vert q_{\mathsf{m+1}}\right\vert
\times \exp \left[ \frac{\beta }{2}\mathsf{m}\right] (\mathsf{m+1)}^{\frac{1%
}{2}}\left\vert q_{\mathsf{m}}\right\vert .
\end{eqnarray*}%
Furthermore, for each $\varepsilon >0$ and every $\mathsf{m}$ we have%
\begin{equation}
\left( \sqrt{\varepsilon }\exp \left[ \frac{\beta }{2}\mathsf{m}\right] (%
\mathsf{m+1)}^{\frac{1}{2}}\left\vert q_{\mathsf{m+1}}\right\vert -\frac{%
\exp \left[ \frac{\beta }{2}\mathsf{m}\right] (\mathsf{m+1)}^{\frac{1}{2}%
}\left\vert q_{\mathsf{m}}\right\vert }{\sqrt{\varepsilon }}\right) ^{2}>0
\label{square}
\end{equation}%
unless%
\begin{equation}
\left\vert q_{\mathsf{m+1}}\right\vert =\frac{\left\vert q_{\mathsf{m}%
}\right\vert }{\varepsilon },  \label{recurrenceter}
\end{equation}%
so that with the exception of this particular case we get 
\begin{eqnarray*}
&&\dsum\limits_{\mathsf{m=0}}^{+\infty }\exp \left[ \beta \mathsf{m}\right] (%
\mathsf{m+1)}\left\vert q_{\mathsf{m+1}}\bar{q}_{\mathsf{m}}\right\vert \\
&<&\frac{1}{2}\left( \varepsilon \dsum\limits_{\mathsf{m=0}}^{+\infty }\exp %
\left[ \beta \mathsf{m}\right] (\mathsf{m+1)}\left\vert q_{\mathsf{m+1}%
}\right\vert ^{2}+\frac{1}{\varepsilon }\dsum\limits_{\mathsf{m=0}}^{+\infty
}\exp \left[ \beta \mathsf{m}\right] (\mathsf{m+1)}\left\vert q_{\mathsf{m}%
}\right\vert ^{2}\right) \\
&=&\frac{1}{2}\left( \varepsilon \exp \left[ -\beta \right] \dsum\limits_{%
\mathsf{m=0}}^{+\infty }\exp \left[ \beta \mathsf{m}\right] \mathsf{m}%
\left\vert q_{\mathsf{m}}\right\vert ^{2}+\frac{1}{\varepsilon }%
\dsum\limits_{\mathsf{m=0}}^{+\infty }\exp \left[ \beta \mathsf{m}\right] (%
\mathsf{m+1)}\left\vert q_{\mathsf{m}}\right\vert ^{2}\right)
\end{eqnarray*}%
by expanding (\ref{square}). Choosing then $\varepsilon =\exp \left[ \beta %
\right] $ and regrouping terms we obtain%
\begin{eqnarray}
&&\dsum\limits_{\mathsf{m=0}}^{+\infty }\exp \left[ \beta \mathsf{m}\right] (%
\mathsf{m+1)}\left\vert q_{\mathsf{m+1}}\bar{q}_{\mathsf{m}}\right\vert 
\notag \\
&<&\frac{1}{2}\left( \dsum\limits_{\mathsf{m=0}}^{+\infty }\exp \left[ \beta 
\mathsf{m}\right] \mathsf{m}\left\vert q_{\mathsf{m}}\right\vert ^{2}+\exp %
\left[ -\beta \right] \dsum\limits_{\mathsf{m=0}}^{+\infty }\exp \left[
\beta \mathsf{m}\right] (\mathsf{m+1)}\left\vert q_{\mathsf{m}}\right\vert
^{2}\right)  \label{estimateter} \\
&=&\frac{1}{2}\dsum\limits_{\mathsf{m=0}}^{+\infty }\exp \left[ \beta 
\mathsf{m}\right] \left( \mathsf{m+}\exp \left[ -\beta \right] \left( 
\mathsf{m+1}\right) \right) \left\vert q_{\mathsf{m}}\right\vert ^{2}, 
\notag
\end{eqnarray}%
which is half of the right-hand side of (\ref{estimatebis}). Similarly, from
(\ref{estimateter}) we have%
\begin{eqnarray}
&&\dsum\limits_{\mathsf{m=0}}^{+\infty }\exp \left[ \beta \left( \mathsf{m-1}%
\right) \right] \mathsf{m}\left\vert q_{\mathsf{m-1}}\bar{q}_{\mathsf{m}%
}\right\vert  \notag \\
&=&\dsum\limits_{\mathsf{m=0}}^{+\infty }\exp \left[ \beta \mathsf{m}\right]
(\mathsf{m+1)}\left\vert q_{\mathsf{m+1}}\bar{q}_{\mathsf{m}}\right\vert
\label{estimatequarto} \\
&<&\frac{1}{2}\dsum\limits_{\mathsf{m=0}}^{+\infty }\exp \left[ \beta 
\mathsf{m}\right] \left( \mathsf{m+}\exp \left[ -\beta \right] \left( 
\mathsf{m+1}\right) \right) \left\vert q_{\mathsf{m}}\right\vert ^{2}  \notag
\end{eqnarray}%
unless (\ref{recurrenceter}) holds, so that the combination of (\ref%
{estimateter}) and (\ref{estimatequarto}) indeed leads to (\ref{estimatebis}%
) up to the exceptional case we alluded to.

Thus, it remains to examine the consequences of the recurrence relation (\ref%
{recurrenceter}), which implies%
\begin{equation*}
\left\vert q_{\mathsf{m}}\right\vert =\exp \left[ -\beta \mathsf{m}\right]
\left\vert q_{\mathsf{0}}\right\vert
\end{equation*}%
for every $\mathsf{m}\in \mathbb{N}$ with our choice of $\varepsilon $. If $%
\mathsf{q}\in D(R^{\ast })$ is real, the preceding expression gives%
\begin{equation*}
q_{\mathsf{m}}=\exp \left[ -\beta \mathsf{m}\right] \left( \pm q_{\mathsf{0}%
}\right) :=\exp \left[ -\beta \mathsf{m}\right] \hat{q}_{\mathsf{0}},
\end{equation*}%
hence (\ref{kernel}) as in Proposition 3 but with some parameter $\hat{q}%
_{0}\in \mathbb{R}$. The general case for $\mathsf{q=r+}i\mathsf{s}$ with $%
\mathsf{r,s}\in D(R^{\ast })$ both real then follows from complexification.
Indeed we have%
\begin{equation*}
\left( R^{\ast }\mathsf{q},\mathsf{q}\right) _{2,w_{\beta }}=\left( R^{\ast }%
\mathsf{r},\mathsf{r}\right) _{2,w_{\beta }}+\left( R^{\ast }\mathsf{s},%
\mathsf{s}\right) _{2,w_{\beta }}+i\left\{ \left( R^{\ast }\mathsf{s},%
\mathsf{r}\right) _{2,w_{\beta }}-\left( R^{\ast }\mathsf{r},\mathsf{s}%
\right) _{2,w_{\beta }}\right\}
\end{equation*}%
where all the inner products are real, a consequence of (\ref{innerproduct})
and (\ref{adjointpremier}) for those in the bracket on the right-hand side.
We then necessarily have%
\begin{equation*}
\left( R^{\ast }\mathsf{q},\mathsf{q}\right) _{2,w_{\beta }}=\left( R^{\ast }%
\mathsf{r},\mathsf{r}\right) _{2,w_{\beta }}+\left( R^{\ast }\mathsf{s},%
\mathsf{s}\right) _{2,w_{\beta }},
\end{equation*}%
so that according to the first part of the proof we have $\left( R^{\ast }%
\mathsf{q},\mathsf{q}\right) _{2,w_{\beta }}<0$ unless both $\mathsf{r}$ and 
$\mathsf{s}$ are of the form%
\begin{equation*}
\mathsf{r=}r_{0}Z_{\beta }\mathsf{p}_{\beta ,\mathsf{Gibbs}}
\end{equation*}%
and 
\begin{equation*}
\mathsf{s=}s_{0}Z_{\beta }\mathsf{p}_{\beta ,\mathsf{Gibbs}}
\end{equation*}%
with $r_{0},s_{0}\in \mathbb{R}$, respectively. The only exceptional case
for which $\left( R^{\ast }\mathsf{q},\mathsf{q}\right) _{2,w_{\beta }}=0$
is therefore $\mathsf{q=}\left( r_{0}+is_{0}\right) Z_{\beta }\mathsf{p}%
_{\beta ,\mathsf{Gibbs}}$.

As for the proof of (\ref{estimate}) we have%
\begin{eqnarray}
&&\left\Vert \left( R^{\ast }-\nu \right) \mathsf{q}\right\Vert _{2,w_{\beta
}}^{2}  \notag \\
&=&\left\Vert R^{\ast }\mathsf{q}\right\Vert _{2,w_{\beta }}^{2}-2\left( 
\func{Re}\nu \right) \left( R^{\ast }\mathsf{q},\mathsf{q}\right)
_{2,w_{\beta }}+\left\vert \nu \right\vert ^{2}\left\Vert \mathsf{q}%
\right\Vert _{2,w_{\beta }}^{2}  \label{quadraticrelation} \\
&\geq &\left\vert \nu \right\vert ^{2}\left\Vert \mathsf{q}\right\Vert
_{2,w_{\beta }}^{2}
\end{eqnarray}%
since $\func{Re}\nu >0$ and $\left( R^{\ast }\mathsf{q},\mathsf{q}\right)
_{2,w_{\beta }}\leqslant 0$ for every $\mathsf{q}\in D(R^{\ast })$, which is
the desired estimate. \ \ $\blacksquare $

\bigskip

The following result and the information already available show that the
entire spectrum of $R^{\ast }$ lies in the interval $\left( -\infty ,0\right]
$:

\bigskip

\textbf{Proposition 4. }\textit{For each }$\nu \in \mathbb{C}$ \textit{such
that} $\func{Re}\nu >0$\textit{\ the inverse operator }$\left( R^{\ast }-\nu
\right) ^{-1}$ \textit{exists on }$l_{\mathbb{C},w_{\beta }}^{2}$\textit{\
and} \textit{we have}%
\begin{equation}
\left\Vert \left( R^{\ast }-\nu \right) ^{-1}\mathsf{p}\right\Vert
_{2,w_{\beta }}\leqslant \left\vert \nu \right\vert ^{-1}\left\Vert \mathsf{p%
}\right\Vert _{2,w_{\beta }}  \label{estimatedecimo}
\end{equation}%
\textit{for every }$\mathsf{p}\in l_{\mathbb{C},w_{\beta }}^{2}$\textit{. In
other words, each such }$\nu $ \textit{belongs to the resolvent set of }$%
R^{\ast }$\textit{.}\bigskip

\textbf{Proof.} It follows from (\ref{estimate}) that there exists at most
one solution to the equation%
\begin{equation*}
\left( R^{\ast }-\nu \right) \mathsf{q=p}
\end{equation*}%
for any $\mathsf{p}\in l_{\mathbb{C},w_{\beta }}^{2}$, namely, that the
operator $R^{\ast }-\nu $ is injective.

Next, we prove that $R^{\ast }-\nu $ is surjective, namely, that 
\begin{equation}
\func{Ran}\left( R^{\ast }-\nu \right) =l_{\mathbb{C},w_{\beta }}^{2}
\label{surjective}
\end{equation}%
where $\func{Ran}\left( R^{\ast }-\nu \right) $ stands for the range of the
operator. This is already true if $\func{Im}\nu \neq 0$ so that it remains
to consider the case $\nu >0.$ Since $R^{\ast }-\nu $ is then self-adjoint
and injective, standard arguments first imply that its range is everywhere
dense in $l_{\mathbb{C},w_{\beta }\text{ }}^{2}$(see, e.g., Chapter X in 
\cite{conway}), hence for every $\mathsf{p}\in l_{\mathbb{C},w_{\beta }}^{2}$
there exists a sequence $\left( \mathsf{q}_{\mathsf{n}}\right) \subset
D(R^{\ast })$ such that $\left( R^{\ast }-\nu \right) \mathsf{q}_{\mathsf{n}%
} $ $\rightarrow \mathsf{p}$ strongly in $l_{\mathbb{C},w_{\beta }}^{2}$ as $%
\mathsf{n\rightarrow +\infty }$. Consequently, (\ref{estimate}) implies that%
\begin{equation*}
\left\Vert \mathsf{q}_{\mathsf{m}}-\mathsf{q}_{\mathsf{n}}\right\Vert
_{2,w_{\beta }}\leq \nu ^{-1}\left\Vert \left( R^{\ast }-\nu \right) \left( 
\mathsf{q}_{\mathsf{m}}-\mathsf{q}_{\mathsf{n}}\right) \right\Vert
_{2,w_{\beta }}\rightarrow 0
\end{equation*}%
as $\mathsf{m},\mathsf{n\rightarrow +\infty }$, hence that $\mathsf{q}_{%
\mathsf{n}}\rightarrow \mathsf{q}$ strongly in $l_{\mathbb{C},w_{\beta
}}^{2} $ as $\mathsf{n\rightarrow +\infty }$ for some $\mathsf{q}$.
Therefore, $\mathsf{q}\in D(R^{\ast })$ and%
\begin{equation*}
\left( R^{\ast }-\nu \right) \mathsf{q=p}
\end{equation*}%
since $R^{\ast }-\nu $ is closed, which gives (\ref{surjective}). The
boundedness of $\left( R^{\ast }-\nu \right) ^{-1}$ for $\nu >0$ then
follows from the inverse mapping theorem, while (\ref{estimatedecimo}) is a
consequence of (\ref{estimate}).\ \ $\blacksquare $

\bigskip

In order to unveil the exact nature of the spectrum of $R^{\ast }$, it is
now essential to provide more concrete information about the domain $%
D(R^{\ast })$. To this end, we consider the vector space $h_{\mathbb{C}%
,w_{\beta }}^{1}$consisting of all complex sequences $\mathsf{p=(}p_{\mathsf{%
m}}\mathsf{)}$ satisfying%
\begin{equation}
\left\Vert \mathsf{p}\right\Vert _{1,2,w_{\beta }}^{2}:=\sum_{\mathsf{m}%
=0}^{+\infty }w_{\beta ,\mathsf{m}}\left( 1+\mathsf{m}^{2}\right) \left\vert
p_{\mathsf{m}}\right\vert ^{2}<+\infty .  \label{normter}
\end{equation}%
We remark that $h_{\mathbb{C},w_{\beta }}^{1}$becomes a separable Hilbert
space in its own right when endowded with the sesquilinear form%
\begin{equation}
\left( \mathsf{p,q}\right) _{1,2,w_{\beta }}:=\sum_{\mathsf{m}=0}^{+\infty
}w_{\beta ,\mathsf{m}}\left( 1+\mathsf{m}^{2}\right) p_{\mathsf{m}}\bar{q}_{%
\mathsf{m}}  \label{sesquilinearform}
\end{equation}%
and the norm (\ref{normter}), but for the time being we consider it as a
vector subspace of $l_{\mathbb{C},w_{\beta }}^{2}$independently of its
intrinsic topology as we wish to show that $D(R^{\ast })$ and $h_{\mathbb{C}%
,w_{\beta }}^{1}$ coincide. We begin with the easy part:

\bigskip

\textbf{Lemma 6. }\textit{We have}%
\begin{equation}
D(R)\subset h_{\mathbb{C},w_{\beta }}^{1}\subseteq D(R^{\ast }),
\label{inclusions}
\end{equation}%
\textit{so that} $h_{\mathbb{C},w_{\beta }}^{1}$\textit{\ is everywhere
dense in }$l_{\mathbb{C},w_{\beta }}^{2}$\textit{. Moreover we have}%
\begin{equation}
R^{\ast }=\bar{R}_{\mathsf{rest}}^{\ast }  \label{equalitysexto}
\end{equation}%
\textit{where }$\bar{R}_{\mathsf{rest}}^{\ast }$\textit{\ stands for the
closure of the restriction }$R_{\mathsf{rest}}^{\ast }$\textit{\ of }$%
R^{\ast }$\textit{\ to }$h_{\mathbb{C},w_{\beta }}^{1}$\textit{.}

\bigskip

\textbf{Proof.} We have $\mathsf{f}_{\mathsf{m}}\in h_{\mathbb{C},w_{\beta
}}^{1}$ for each $\mathsf{m}$ and therefore $D(R)\subseteq $ $h_{\mathbb{C}%
,w_{\beta }}^{1}$. Moreover $\mathsf{p}_{\beta ,\mathsf{Gibbs}}\in h_{%
\mathbb{C},w_{\beta }}^{1}$, so that the strict inclusion in (\ref%
{inclusions}) holds since $\mathsf{p}_{\beta ,\mathsf{Gibbs}}\notin D(R)$.
Furthermore, $h_{\mathbb{C},w_{\beta }}^{1}$ is everywhere dense in $l_{%
\mathbb{C},w_{\beta }}^{2}$ since $D(R)$ is. Now let $\mathsf{q}\in h_{%
\mathbb{C},w_{\beta }}^{1}$ be arbitrary and let us define $\mathsf{q}^{\ast
}$ by 
\begin{equation*}
q_{\mathsf{m}}^{\ast }:=\rho \left\{ \left( \mathsf{m}+1\right) q_{\mathsf{%
m+1}}+\exp \left[ -\beta \right] \mathsf{m}q_{\mathsf{m-1}}-(\mathsf{m}+\exp %
\left[ -\beta \right] (\mathsf{m}+1))q_{\mathsf{m}}\right\}
\end{equation*}%
for every $\mathsf{m}\in \mathbb{N}$, having (\ref{adjointpremier}) in mind.
From this and (\ref{normter}) we easily obtain the estimate%
\begin{equation}
\left\Vert \mathsf{q}^{\ast }\right\Vert _{2,w_{\beta }}\leq c_{\beta ,\rho
}\left\Vert \mathsf{q}\right\Vert _{1,2,w_{\beta }}<+\infty
\label{estimateoitavo}
\end{equation}%
where $c_{\beta ,\rho }>0$ is a positive constant depending only on $\beta $
and $\rho $, which means that $\mathsf{q}^{\ast }\in l_{\mathbb{C},w_{\beta
}}^{2}$. The fact that%
\begin{equation*}
\left( R\mathsf{p},\mathsf{q}\right) _{2,w_{\beta }}=\left( \mathsf{p},%
\mathsf{q}^{\ast }\right) _{2,w_{\beta }}
\end{equation*}%
for every $\mathsf{p}\in D(R)$ then follows from a computation similar to
that carried out in the proof of Lemma 4, which proves the second inclusion
in (\ref{inclusions}).

Finally, writing $R_{\mathsf{rest}}^{\ast }$ for the restriction of $R^{\ast
}$ to $h_{\mathbb{C},w_{\beta }}^{1}$, which is symmetric and closable
according the the second inclusion in (\ref{inclusions}), we have%
\begin{equation*}
R\subseteq R_{\mathsf{rest}}^{\ast }\subseteq R^{\ast }
\end{equation*}%
and consequently%
\begin{equation*}
R^{\ast }\supseteq \left( R_{\mathsf{rest}}^{\ast }\right) ^{\ast }\supseteq
R^{\ast }
\end{equation*}%
since $R^{\ast }$ is self-adjoint, which means that $R^{\ast }=$ $\left( R_{%
\mathsf{rest}}^{\ast }\right) ^{\ast }$ and thereby%
\begin{equation*}
R^{\ast }=R^{\ast \ast }=\left( R_{\mathsf{rest}}^{\ast }\right) ^{\ast \ast
}=\bar{R}_{\mathsf{rest}}^{\ast }
\end{equation*}%
where $\bar{R}_{\mathsf{rest}}^{\ast }$ stands for the closure of $R_{%
\mathsf{rest}}^{\ast }$. \ \ $\blacksquare $

\bigskip

Let us now decompose $R_{\mathsf{rest}}^{\ast }$ into a sum of two
operators, namely,%
\begin{equation}
R_{\mathsf{rest}}^{\ast }=S+T  \label{decomposition}
\end{equation}%
where%
\begin{eqnarray}
\left( S\mathsf{q}\right) _{\mathsf{m}} &:&=-\rho (\mathsf{m}+\exp \left[
-\beta \right] (\mathsf{m}+1))q_{\mathsf{m}},  \notag \\
\left( T\mathsf{q}\right) _{\mathsf{m}} &:&=\rho \left\{ \left( \mathsf{m}%
+1\right) q_{\mathsf{m+1}}+\exp \left[ -\beta \right] \mathsf{m}q_{\mathsf{%
m-1}}\right\}  \label{deomposition}
\end{eqnarray}%
for every $\mathsf{m}\in \mathbb{N}$, both defined on $h_{\mathbb{C}%
,w_{\beta }}^{1}.$ It is easily verified that $S$ is a diagonal operator
with respect to the basis $\left( \mathsf{f}_{\mathsf{m}}\right) $, while $T$
is the sum of an upper and a lower triangular operator. The following result
holds:

\bigskip

\textbf{Lemma 7. }\textit{Let }$\nu >0$\textit{\ and let us define }$S_{\nu
}:=S-\nu $\textit{. Then }$S_{\nu }$ \textit{is self-adjoint on} $h_{\mathbb{%
C},w_{\beta }}^{1}$ \textit{and its inverse }$S_{\nu }^{-1}$\textit{\ exists
and is compact in }$l_{\mathbb{C},w_{\beta }}^{2}$\textit{. Moreover, the
operator} $TS_{\nu }^{-1}$\textit{\ remains bounded}.

\bigskip

\textbf{Proof. }We first prove that $S$ is self-adjoint on $h_{\mathbb{C}%
,w_{\beta }}^{1}$, namely, that%
\begin{equation}
h_{\mathbb{C},w_{\beta }}^{1}\subseteq D(S^{\ast })\subseteq h_{\mathbb{C}%
,w_{\beta }}^{1}.  \label{inclusionbis}
\end{equation}%
Indeed $S$ is obviously symmetric with respect to (\ref{innerproduct}), with%
\begin{equation*}
\left( S^{\ast }\mathsf{q}\right) _{\mathsf{m}}=-\rho (\mathsf{m}+\exp \left[
-\beta \right] (\mathsf{m}+1))q_{\mathsf{m}}
\end{equation*}%
for every $\mathsf{m}$, so that the first inclusion in (\ref{inclusionbis})
holds. In order to prove the second inclusion, we remark that if\textsf{\ }$%
\mathsf{q}\in D(S^{\ast })$ then from the very definition of this domain we
have%
\begin{equation*}
\dsum\limits_{\mathsf{m}=0}^{+\infty }\exp \left[ \beta \mathsf{m}\right]
\left( \mathsf{m+}\exp \left[ -\beta \right] \mathsf{(m+1)}\right)
^{2}\left\vert q_{\mathsf{m}}\right\vert ^{2}<+\infty ,
\end{equation*}%
from which the estimate%
\begin{equation*}
\left\Vert \mathsf{q}\right\Vert _{1,2,w_{\beta }}^{2}\leq \tilde{c}_{\beta
,\rho }\left\Vert S^{\ast }\mathsf{q}\right\Vert _{2,w_{\beta }}^{2}<+\infty
\end{equation*}%
follows with some $\tilde{c}_{\beta ,\rho }>0$ so that $\mathsf{q}\in h_{%
\mathbb{C},w_{\beta }}^{1}$. Therefore, $S$ is self-adjoint on $h_{\mathbb{C}%
,w_{\beta }}^{1}$ and so is $S_{\nu }$.

Regarding the existence and the compactness of $S_{\nu }^{-1}$ we first note
that%
\begin{equation}
\left( S_{\nu }^{-1}\mathsf{q}\right) _{\mathsf{m}}=a_{\mathsf{m}}q_{\mathsf{%
m}}  \label{inverse}
\end{equation}%
for each $\mathsf{q}\in $ $l_{\mathbb{C},w_{\beta }}^{2}$ and every $\mathsf{%
m}$, where%
\begin{equation}
a_{\mathsf{m}}\mathsf{\ :=-}\left( \rho (\mathsf{m}+\nu _{\rho }+\exp \left[
-\beta \right] (\mathsf{m}+1))\right) ^{-1}  \label{sequence}
\end{equation}%
with $\nu _{\rho }:=\rho ^{-1}\nu >0$. Then $a_{\mathsf{m}}\rightarrow 0$ as 
$\mathsf{m\rightarrow +\infty }$, so that the desired compactness follows
from the fact that $S_{\nu }^{-1}$ may be viewed as the uniform limit of the
sequence of finite-rank operators $A_{\mathsf{N}}$ given by%
\begin{equation*}
\left( A_{_{\mathsf{N}}}\mathsf{q}\right) _{\mathsf{m}}=\left\{ 
\begin{array}{c}
a_{\mathsf{m}}q_{\mathsf{m}}\text{ for }0\leq \mathsf{m\leq N,} \\ 
\\ 
0\text{ \ \ \ \ \ for }\mathsf{m\geq N+1.}%
\end{array}%
\right.
\end{equation*}%
As for the last statement of the lemma, we claim that%
\begin{equation}
\left\Vert TS_{\nu }^{-1}\mathsf{q}\right\Vert _{2,w_{\beta }}\leq c_{\beta
}\left\Vert \mathsf{q}\right\Vert _{2,w_{\beta }}  \label{boundedoperator}
\end{equation}%
for every $\mathsf{q}\in l_{\mathbb{C},w_{\beta }}^{2}$ and some $c_{\beta
}>0$ depending solely on $\beta $. Indeed we have%
\begin{equation}
\left( TS_{\nu }^{-1}\mathsf{q}\right) _{\mathsf{m}}=\rho \left\{ \left( 
\mathsf{m}+1\right) \left( S_{\nu }^{-1}\mathsf{q}\right) _{\mathsf{m+1}%
}+\exp \left[ -\beta \right] \mathsf{m}\left( S_{\nu }^{-1}\mathsf{q}\right)
_{\mathsf{m-1}}\right\}  \label{component}
\end{equation}%
for every $\mathsf{m}$ from the definition of $T$ in (\ref{deomposition}),
along with the estimates%
\begin{equation}
\left\vert \left( S_{\nu }^{-1}\mathsf{q}\right) _{\mathsf{m+1}}\right\vert
=\left\vert a_{\mathsf{m+1}}q_{\mathsf{m+1}}\right\vert \leq \frac{%
\left\vert q_{\mathsf{m+1}}\right\vert }{\rho \left( \mathsf{m}+1\right) }
\label{estimatenono}
\end{equation}%
and%
\begin{equation*}
\left\vert \left( S_{\nu }^{-1}\mathsf{q}\right) _{\mathsf{m-1}}\right\vert
=\left\vert a_{\mathsf{m-1}}q_{\mathsf{m-1}}\right\vert \leq \frac{\exp %
\left[ \beta \right] \left\vert q_{\mathsf{m-1}}\right\vert }{\rho \mathsf{m}%
}
\end{equation*}%
from (\ref{sequence}), with $\mathsf{m\neq 0}$ in the last inequality.
Therefore we obtain%
\begin{equation*}
\left\vert \left( TS_{\nu }^{-1}\mathsf{q}\right) _{\mathsf{m}}\right\vert
^{2}\leq 2\left( \left\vert q_{\mathsf{m+1}}\right\vert ^{2}+\left\vert q_{%
\mathsf{m-1}}\right\vert ^{2}\right)
\end{equation*}%
for every $\mathsf{m\neq 0}$, while for $\mathsf{m}=0$ we have%
\begin{equation*}
\left\vert \left( TS_{\nu }^{-1}\mathsf{q}\right) _{\mathsf{0}}\right\vert
^{2}\leq \left\vert q_{\mathsf{1}}\right\vert ^{2}
\end{equation*}%
from (\ref{component}) and (\ref{estimatenono}). Lumping all things together
using (\ref{norm}) we then get%
\begin{equation*}
\left\Vert TS_{\nu }^{-1}\mathsf{q}\right\Vert _{2,w_{\beta }}^{2}\leq 
\tilde{c}_{\beta }\left\Vert \mathsf{q}\right\Vert _{2,w_{\beta }}^{2}
\end{equation*}%
for some $\tilde{c}_{\beta }>0$, which gives the desired estimate. \ \ $%
\blacksquare $

\bigskip

\textsc{Remark.} Relation (\ref{boundedoperator}) becomes 
\begin{equation*}
\left\Vert T\mathsf{p}\right\Vert _{2,w_{\beta }}\leq c_{\beta }\left\Vert
S_{\nu }\mathsf{p}\right\Vert _{2,w_{\beta }}
\end{equation*}%
with $\mathsf{p}=S_{\nu }^{-1}\mathsf{q}$, which shows that the symmetric
operator $T$ is relatively bounded with respect to $S_{\nu }$. However, the
above bound is not one that allows us to apply directly the Kato-Rellich
theory to prove the self-adjointness of $S+T$ on $h_{\mathbb{C},w_{\beta
}}^{1}$ (see, e.g., Section 4 in Chapter V of \cite{Kato}).

\bigskip

In order to determine the ultimate structure of the spectrum of $R^{\ast }$
we still need, therefore, the following auxiliary result:

\bigskip

\textbf{Proposition 5.} \textit{We have }$D(R^{\ast })=h_{\mathbb{C}%
,w_{\beta }}^{1}$,\textit{\ and on this domain the operator }$R^{\ast }$%
\textit{\ has a compact resolvent}.

\bigskip

\textbf{Proof. }For every $\mathsf{q\in }$ $h_{\mathbb{C},w_{\beta }}^{1}$
we may write%
\begin{equation}
\left( R^{\ast }-\nu \right) \mathsf{q}=\left( S_{\nu }+T\right) \mathsf{q}%
=\left( I+TS_{\nu }^{-1}\right) S_{\nu }\mathsf{q}  \label{decopositionbis}
\end{equation}%
from the decomposition (\ref{decomposition}) since $R^{\ast }=$ $R_{\mathsf{%
rest}}^{\ast }$ on that subspace, where $I$ stands for the identity operator
and $\nu >0$. Equivalently we have%
\begin{equation*}
\left( R^{\ast }-\nu \right) S_{\nu }^{-1}\mathsf{p}=\left( I+TS_{\nu
}^{-1}\right) \mathsf{p}
\end{equation*}%
whenever $\mathsf{p}=S_{\nu }\mathsf{q}$, which implies the equality of the
two operators%
\begin{equation}
\left( R^{\ast }-\nu \right) S_{\nu }^{-1}=I+TS_{\nu }^{-1}  \label{operator}
\end{equation}%
on $\func{Ran}S_{\nu }=D\left( S_{\nu }^{-1}\right) =l_{\mathbb{C},w_{\beta
}}^{2}$.

We now proceed by showing that the operator on the right-hand side of (\ref%
{operator}) has a bounded inverse. On the one hand, the equation%
\begin{equation*}
\left( R^{\ast }-\nu \right) S_{\nu }^{-1}\mathsf{q=0}
\end{equation*}%
with $\mathsf{q}\in l_{\mathbb{C},w_{\beta }}^{2}$ implies $S_{\nu }^{-1}%
\mathsf{q=0}$ as a consequence of (\ref{estimate}) and thereby $\mathsf{q=0}$
from (\ref{inverse}), so that the operator in (\ref{operator}) is injective.
On the other hand, for any given $\mathsf{p}\in l_{\mathbb{C},w_{\beta
}}^{2} $ the equation%
\begin{equation*}
\left( R^{\ast }-\nu \right) S_{\nu }^{-1}\mathsf{q=p}
\end{equation*}%
implies

\begin{equation*}
S_{\nu }^{-1}\mathsf{q}=\left( R^{\ast }-\nu \right) ^{-1}\mathsf{p}\in h_{%
\mathbb{C},w_{\beta }}^{1}
\end{equation*}%
and so has the unique solution 
\begin{equation*}
\mathsf{q=S_{\nu }\left( R^{\ast }-\nu \right) }^{-1}\mathsf{p,}
\end{equation*}%
so that the operator in (\ref{operator}) is also surjective. Therefore, from
the inverse mapping theorem we infer that the operator on the right-hand
side of (\ref{operator}) has a bounded inverse in $l_{\mathbb{C},w_{\beta
}}^{2}$, as desired.

Since $\left( I+TS_{\nu }^{-1}\right) ^{-1}$ is bounded, we now easily infer
that the operator%
\begin{equation*}
R_{\mathsf{rest}}^{\ast }-\nu =\left( I+TS_{\nu }^{-1}\right) S_{\nu }
\end{equation*}%
is closed since both operators $S_{\nu }$ and $I+TS_{\nu }^{-1}$ are closed
(see, e.g., Section 5 in Chapter III\ of \cite{Kato}). Consequently $R_{%
\mathsf{rest}}^{\ast }$ is closed as well so that%
\begin{equation*}
R^{\ast }=\bar{R}_{\mathsf{rest}}^{\ast }=R_{\mathsf{rest}}^{\ast }
\end{equation*}%
from (\ref{equalitysexto}), which proves the desired equality $D(R^{\ast
})=h_{\mathbb{C},w_{\beta }}^{1}$.

We complete the proof of the proposition by showing that the operator $%
R^{\ast }$ has a compact resolvent. From (\ref{decopositionbis}) we first
infer that the operator equality%
\begin{equation*}
\left( R^{\ast }-\nu \right) ^{-1}=S_{\nu }^{-1}\left( I+TS_{\nu
}^{-1}\right) ^{-1}
\end{equation*}%
holds on $h_{\mathbb{C},w_{\beta }}^{1}$, and hence on $l_{\mathbb{C}%
,w_{\beta }}^{2}$ since $h_{\mathbb{C},w_{\beta }}^{1}$ is everywhere dense
therein. Therefore $\left( R^{\ast }-\nu \right) ^{-1}$ is compact as the
product of a compact operator with a bounded operator. \ \ $\blacksquare $

\bigskip

\textsc{Remarks.} (1) We can easily verify that the compactness of the
resolvent of $R^{\ast }$ implies the compactness of the embedding%
\begin{equation}
h_{\mathbb{C},w_{\beta }}^{1}\hookrightarrow l_{\mathbb{C},w_{\beta }}^{2}
\label{embeddingter}
\end{equation}%
when we consider $h_{\mathbb{C},w_{\beta }}^{1}$ as endowed with the
sesquilinear form (\ref{sesquilinearform}) and the induced norm (\ref%
{normter}). Indeed, for each $\nu >0$ and every $\mathsf{q}\in h_{\mathbb{C}%
,w_{\beta }}^{1}$ we have the inequality%
\begin{equation*}
\left\Vert \left( R^{\ast }-\nu \right) \mathsf{q}\right\Vert _{2,w_{\beta
}}\leq c_{\beta ,\rho ,\nu }\left\Vert \mathsf{q}\right\Vert _{1,2,w_{\beta
}}
\end{equation*}%
for some $c_{\beta ,\rho ,\nu }>0$ as a consequence of (\ref{normter}) and (%
\ref{estimateoitavo}). Therefore $R^{\ast }-\nu $ is a linear bounded
operator from $h_{\mathbb{C},w_{\beta }}^{1}$ into $l_{\mathbb{C},w_{\beta
}}^{2}$, which implies that%
\begin{equation*}
i_{c}:=\left( R^{\ast }-\nu \right) ^{-1}\left( R^{\ast }-\nu \right)
\end{equation*}%
is compact since $\left( R^{\ast }-\nu \right) ^{-1}$ is compact. But $i_{c}$
is the restriction of the identity operator $I$ to $h_{\mathbb{C},w_{\beta
}}^{1}$, that is, the embedding map. In the appendix we shall derive the
compactness of (\ref{embeddingter}) independently from a more general
embedding result which involves an intermediary space between $h_{\mathbb{C}%
,w_{\beta }}^{1}$ and $l_{\mathbb{C},w_{\beta }}^{2}$.

(2) The compactness of (\ref{embeddingter}) also allows us to prove the
compactness of the operator $S_{\nu }^{-1}$ in a different manner, by
showing that the set%
\begin{equation}
D_{\theta }=\left\{ \mathsf{q}\in h_{\mathbb{C},w_{\beta }}^{1}:\left\Vert 
\mathsf{q}\right\Vert _{2,w_{\beta }}\leq 1,\left\Vert S_{\nu }\mathsf{q}%
\right\Vert _{2,w_{\beta }}\leq \theta \right\}  \label{set}
\end{equation}%
where $\theta \in \left[ 0,+\infty \right) $ is compact in $l_{\mathbb{C}%
,w_{\beta }}^{2}$. For the sake of completeness we shall provide the details
in the appendix.

\bigskip

We can now state and prove the main result of this section:

\bigskip

\textbf{Theorem.} \textit{The following statements hold:}

\textit{(a) The operator }$R^{\ast }$ \textit{has a purely discrete
spectrum. More specifically, there exists an orthonormal basis }$\left( 
\mathsf{\hat{p}}_{\mathsf{k}}\right) _{\mathsf{k}\in \mathbb{N}}$ \textit{of 
}$l_{\mathbb{C},w_{\beta }}^{2}$ \textit{consisting exclusively of
eigenvectors of }$R^{\ast }$, \textit{with }$\mathsf{\hat{p}}_{\mathsf{k}%
}\in h_{\mathbb{C},w_{\beta }}^{1}$\textit{\ satisfying}%
\begin{equation}
R^{\ast }\mathsf{\hat{p}}_{\mathsf{k}}=\nu _{\mathsf{k}}\mathsf{\hat{p}}_{%
\mathsf{k}}  \label{eigenequationbis}
\end{equation}%
\textit{for each }$\mathsf{k}$,\textit{\ the corresponding eigenvalues being
such that}%
\begin{equation}
\dim _{\mathbb{C}}\func{Ker}\left( R^{\ast }-\nu _{\mathsf{k}}\right) =1
\label{dimension}
\end{equation}%
\textit{and which we order as}%
\begin{equation}
\nu _{\mathsf{k+1}}<\nu _{\mathsf{k}}<\nu _{0}:=0  \label{ordering}
\end{equation}%
\textit{for every} $\mathsf{k}\in \left\{ 1,2,...\right\} $\textit{.} 
\textit{Moreover,} $\nu _{\mathsf{k}}\rightarrow -\infty $ \textit{as} $%
\mathsf{k\rightarrow +\infty }$.

\bigskip \textit{(b) The operator }$R^{\ast }$\textit{generates a
holomorphic semigroup of contractions\ in }$l_{\mathbb{C},w_{\beta }}^{2}$, 
\textit{written} $\exp \left[ tR^{\ast }\right] $ \textit{for} $t\in \left[
0,+\infty \right) $, \textit{and the norm-convergent spectral resolution}%
\begin{equation}
\exp \left[ tR^{\ast }\right] \mathsf{p=}\dsum\limits_{\mathsf{k}%
=0}^{+\infty }\left( \mathsf{p,\hat{p}}_{\mathsf{k}}\right) _{_{2,w_{\beta
}}}\exp \left[ t\nu _{\mathsf{k}}\right] \mathsf{\hat{p}}_{\mathsf{k}}
\label{spectralresolution}
\end{equation}%
\textit{holds} \textit{for each} $\mathsf{p}\in l_{\mathbb{C},w_{\beta
}}^{2} $ \textit{and every }$t\in \left[ 0,+\infty \right) $. \textit{%
Moreover, this semigroup is self-adjoint in} $l_{\mathbb{C},w_{\beta }}^{2}$.

\bigskip

\textbf{Proof. }The first part as well as the very last part of Statement
(a) are an immediate consequence of Lemma 5 and Proposition 5 (see, e.g.,
Section 3 in Chapter IX of \cite{edmundsevans} on the relation between
operators with compact resolvent and spectra). As for the non-degeneracy of
the eigenvalues, we note that the equation%
\begin{equation*}
\left( R^{\ast }-\nu _{\mathsf{k}}\right) \mathsf{\hat{p}}_{\mathsf{k}}=0
\end{equation*}%
is equivalent to the recurrence relation%
\begin{equation*}
\hat{p}_{\mathsf{k},\mathsf{m}+1}=\frac{\left( \mathsf{m}+\nu _{\mathsf{k,}%
\rho }+\exp \left[ -\beta \right] (\mathsf{m}+1)\right) \hat{p}_{\mathsf{k},%
\mathsf{m}}-\exp \left[ -\beta \right] \mathsf{m}\hat{p}_{\mathsf{k},\mathsf{%
m}-1}}{\mathsf{m+1}}
\end{equation*}%
for the components of the eigenvectors, where $\nu _{\mathsf{k,}\rho }:=\rho
^{-1}\nu _{\mathsf{k}}$. Therefore%
\begin{eqnarray*}
\hat{p}_{\mathsf{k},1} &=&\left( \nu _{\mathsf{k,}\rho }+\exp \left[ -\beta %
\right] \right) \hat{p}_{\mathsf{k},\mathsf{0}}, \\
\hat{p}_{\mathsf{k},2} &=&\frac{\left( \mathsf{1}+\nu _{\mathsf{k,}\rho
}+2\exp \left[ -\beta \right] \right) \hat{p}_{\mathsf{k},\mathsf{1}}-\exp %
\left[ -\beta \right] \hat{p}_{\mathsf{k},\mathsf{0}}}{\mathsf{2}}
\end{eqnarray*}%
and so on, each component being ultimately of the form%
\begin{equation*}
\hat{p}_{\mathsf{k},\mathsf{m}}=q_{\mathsf{m,}\beta ,\nu _{\mathsf{k,}\rho }}%
\hat{p}_{\mathsf{k},\mathsf{0}}
\end{equation*}%
where the constants $q_{\mathsf{m,}\beta ,\nu _{\mathsf{k,}\rho }}\in 
\mathbb{C}$ are fixed and depend solely on $\mathsf{m},\beta $ and $\nu _{%
\mathsf{k,}\rho }$, whereas $\hat{p}_{\mathsf{k},\mathsf{0}}\in \mathbb{C}$
is a free parameter. Therefore Relation (\ref{dimension}) holds.

As for the proof of statement (b), Proposition 4 implies that the hypotheses
of the Hille-Yosida theorem hold true, so that $R^{\ast }$ indeed generates
a semigroup of contractions\ in\textit{\ }$l_{\mathbb{C},w_{\beta }}^{2}$%
which we denote by $\exp \left[ tR^{\ast }\right] $ for $t\in \left[
0,+\infty \right) $. Furthermore, for a fixed $\varepsilon >0$ let us define
the auxiliary semigroup%
\begin{equation}
S(t):=\exp \left[ -\varepsilon t\right] \exp \left[ tR^{\ast }\right]
\label{semigroup}
\end{equation}%
whose infinitesimal generator is $R_{\varepsilon }^{\ast }:=R^{\ast
}-\varepsilon $. We then infer from Proposition 4 that $\left( R^{\ast
}-\varepsilon \right) ^{-1}$ exists as a linear bounded operator on $l_{%
\mathbb{C},w_{\beta }}^{2}$, hence that $\nu =0$ belongs to the resolvent
set of $R_{\varepsilon }^{\ast }$. Moreover, for every $\nu \in \mathbb{C}$
with $\func{Re}\nu >0$, $\func{Im}\nu \neq 0$, and taking the inequality $%
\left( R_{\varepsilon }^{\ast }\mathsf{q},\mathsf{q}\right) _{2,w_{\beta
}}\leq 0$ into account, an estimate similar to (\ref{estimate}) gives%
\begin{equation*}
\left\Vert \left( R_{\varepsilon }^{\ast }-\nu \right) \mathsf{q}\right\Vert
_{2,w_{\beta }}\geq \left\vert \nu \right\vert \left\Vert \mathsf{q}%
\right\Vert _{2,w_{\beta }}\geq \left\vert \func{Im}\nu \right\vert
\left\Vert \mathsf{q}\right\Vert _{2,w_{\beta }},
\end{equation*}%
so that the supremum norm of $\left( R_{\varepsilon }^{\ast }-\nu \right)
^{-1}$ satisfies%
\begin{equation*}
\left\Vert \left( R_{\varepsilon }^{\ast }-\nu \right) ^{-1}\right\Vert
_{\infty }\leq \left\vert \func{Im}\nu \right\vert ^{-1}.
\end{equation*}%
Therefore, the semigroup generated by $R_{\varepsilon }^{\ast }$ is
holomorphic (see, e.g., Condition (b) of Theorem 5.2 in \cite{pazy}), and so
is $\exp \left[ tR^{\ast }\right] $ according to (\ref{semigroup}). Now, for
any $\mathsf{p}\in l_{\mathbb{C},w_{\beta }}^{2}$ we have the
norm-convergent expansion%
\begin{equation*}
\mathsf{p=}\dsum\limits_{\mathsf{k}=0}^{+\infty }\left( \mathsf{p,\hat{p}}_{%
\mathsf{k}}\right) _{2,w_{\beta }}\mathsf{\hat{p}}_{\mathsf{k}},
\end{equation*}%
so that (\ref{spectralresolution}) follows from the standard spectral theory
of semigroups. Finally, from (\ref{spectralresolution}) we obtain%
\begin{equation*}
\left( \exp \left[ tR^{\ast }\right] \mathsf{p,q}\right) _{2,w_{\beta
}}=\left( \mathsf{p},\exp \left[ tR^{\ast }\right] \mathsf{q}\right)
_{2,w_{\beta }}
\end{equation*}%
for all $\mathsf{p,q\in }$\ $l_{\mathbb{C},w_{\beta }}^{2}$ at once. \ \ $%
\blacksquare $

\bigskip

An immediate consequence of the preceding result is the following:

\bigskip

\textbf{Corollary 1.} \textit{Let }$\mathsf{p}^{\ast }\in l_{\mathbb{C}%
,w_{\beta }}^{2}$ \textit{satisfy the two conditions in (\ref{probabilities}%
), and let us define}%
\begin{equation}
\mathsf{p}\left( t\right) :=\exp \left[ tR^{\ast }\right] \mathsf{p}^{\ast }%
\mathsf{\label{evolution}}
\end{equation}%
\textit{for every} $t\in \left[ 0,+\infty \right) $. \textit{Assuming that
the eigenvalues of }$R^{\ast }$\textit{are ordered as in (\ref{ordering}) we
then have}%
\begin{equation}
\left\Vert \mathsf{p}\left( t\right) -\mathsf{p}_{\beta ,\mathsf{Gibbs}%
}\right\Vert _{2,w_{\beta }}\leq \exp \left[ -t\left\vert \nu
_{1}\right\vert \right] \left\Vert \mathsf{p}^{\ast }\right\Vert
_{2,w_{\beta }}  \label{expdecay}
\end{equation}%
\textit{for each }$t\in \left[ 0,+\infty \right) $. \textit{Moreover, the
function }$t\rightarrow \mathsf{p}\left( t\right) =(p_{\mathsf{m}}(t))$ 
\textit{is differentiable for every }$t\in \left( 0,+\infty \right) $ 
\textit{and we have}%
\begin{equation}
\frac{d\mathsf{p}\left( t\right) }{dt}=R^{\ast }\mathsf{p}\left( t\right)
\label{equation}
\end{equation}%
\textit{with}%
\begin{equation}
\left\Vert \frac{d\mathsf{p}\left( t\right) }{dt}\right\Vert _{2,w_{\beta
}}\leq c_{\beta ,\rho }t^{-1}\left\Vert \mathsf{p}^{\ast }\right\Vert
_{2,w_{\beta }}  \label{otherdecay}
\end{equation}%
\textit{for some constant }$c_{\beta ,\rho }>0$. \textit{Finally,}%
\begin{equation}
p_{_{\mathsf{m}}}\left( t\right) \geq 0\text{, \ \ \ }\sum_{\mathsf{m=0}%
}^{+\infty }p_{_{\mathsf{m}}}\left( t\right) =1  \label{probabilitybis}
\end{equation}%
\textit{for every} $t\in \left[ 0,+\infty \right) $.

\bigskip

\textbf{Proof. }From (\ref{spectralresolution}) we have%
\begin{eqnarray*}
&&\left\Vert \exp \left[ tR^{\ast }\right] \mathsf{p}^{\ast }\mathsf{-\left( 
\mathsf{p}^{\ast }\mathsf{,\hat{p}}_{\mathsf{0}}\right) }_{2,w_{\beta }}%
\mathsf{\hat{p}_{\mathsf{0}}}\right\Vert _{2,w_{\beta }}^{2} \\
&\mathsf{=}&\dsum\limits_{\mathsf{k}=1}^{+\infty }\left\vert \left( \mathsf{p%
}^{\ast }\mathsf{,\hat{p}}_{\mathsf{k}}\right) _{2,w_{\beta }}\right\vert
^{2}\exp \left[ -2t\left\vert \nu _{\mathsf{k}}\right\vert \right] \leq \exp %
\left[ -2t\left\vert \nu _{\mathsf{1}}\right\vert \right] \left\Vert \mathsf{%
p}^{\ast }\right\Vert _{2,w_{\beta }}^{2}
\end{eqnarray*}%
according to (\ref{ordering}), so that (\ref{expdecay}) will follow if we
can prove the relation%
\begin{equation}
\left( \mathsf{p}^{\ast }\mathsf{,\hat{p}}_{\mathsf{0}}\right) _{2,w_{\beta
}}\mathsf{\hat{p}}_{\mathsf{0}}=\mathsf{p}_{\beta ,\mathsf{Gibbs}}.
\label{relationbis}
\end{equation}%
As $\mathsf{\hat{p}}_{\mathsf{0}}$ is an eigenvector of $R^{\ast }$
corresponding to the eigenvalue $\nu _{0}=0$ and satisfying $\left\Vert 
\mathsf{\hat{p}}_{\mathsf{0}}\right\Vert _{2,w_{\beta }}=1$, we may choose $%
\mathsf{\hat{p}}_{\mathsf{0}}=Z_{\beta }^{\frac{1}{2}}\mathsf{p}_{\beta ,%
\mathsf{Gibbs}}$ without restricting the generality where $Z_{\beta }$ is
given by (\ref{partittionfunction}). Therefore we obtain%
\begin{equation*}
\left( \mathsf{p}^{\ast }\mathsf{,\hat{p}}_{\mathsf{0}}\right) _{2,w_{\beta
}}=Z_{\beta }^{-\frac{1}{2}}
\end{equation*}%
as a consequence of the second relation in (\ref{probabilities}), which
leads to (\ref{relationbis}).

Next, as a holomorphic semigroup, $\exp \left[ tR^{\ast }\right] _{t\in %
\left[ 0,+\infty \right) }$ is \textit{a fortiori }differentiable and
equation (\ref{equation}) holds. Furthermore we have%
\begin{equation*}
\left\Vert R^{\ast }\exp \left[ tR^{\ast }\right] \right\Vert _{\infty }\leq
c_{\beta ,\rho }t^{-1}
\end{equation*}%
for some\textit{\ }$c_{\beta ,\rho }>0$ and every $t\in \left( 0,+\infty
\right) $ for the uniform norm of the operator as a characteristic property
of holomorphic semigroups (see, e.g, Section 2.5 in Chapter 2 of \cite{pazy}%
), which proves (\ref{otherdecay}).

As for the proof of the first relation in (\ref{probabilitybis}), let us
define the family of orthogonal projections $\left( Q_{\mathsf{m}}\right) _{%
\mathsf{m}\in \mathbb{N}}$ in $l_{\mathbb{C},w_{\beta }}^{2}$ by $\left( Q_{%
\mathsf{m}}\mathsf{q}\right) _{\mathsf{k}}=q_{\mathsf{m}}\delta _{\mathsf{m,k%
}}$ for each $\mathsf{q}\in $ $l_{\mathbb{C},w_{\beta }}^{2}$ and every $%
\mathsf{k}\in \mathbb{N}$. Then we have 
\begin{equation}
p_{\mathsf{m}}(t)=Q_{\mathsf{m}}\exp \left[ tR^{\ast }\right] \mathsf{p}%
^{\ast }  \label{projection}
\end{equation}%
for every $t\in $ $\left[ 0,+\infty \right) $, and from the semigroup
property we see that it is sufficient to have $p_{\mathsf{m}}(t)\geq 0$ for
all sufficiently small $t\geq 0$ in order that the inequality be valid for
every $t$. Now, the function given in (\ref{projection}) is continuous for $%
t>0$ and right-continuous at $t=0$ for every $\mathsf{m}$. Therefore, it is
impossible to have an $\mathsf{m}^{\ast }\in \mathbb{N}$ and a sufficiently
small $t^{\ast }>0$ such that $p_{\mathsf{m}^{\ast }}(t^{\ast })<0$, for by
hypothesis either $p_{\mathsf{m}^{\ast }}(0)>0$ or $p_{\mathsf{m}^{\ast
}}(0)=0$. In the first case, $p_{\mathsf{m}^{\ast }}(t^{\ast })<0$ would
contradict the right-continuity of (\ref{projection}) for $\mathsf{m=m}%
^{\ast }$ at $t=0$. In the second case, equation (\ref{masterequationster})
for $\mathsf{m=m}^{\ast }$ gives%
\begin{equation*}
\lim_{t\searrow 0_{+}}\frac{dp_{\mathsf{0}}(t)}{dt}=\rho p_{\mathsf{1}%
}(0)\geq 0
\end{equation*}%
if $\mathsf{m}^{\ast }=0$, and

\begin{equation*}
\lim_{t\searrow 0_{+}}\frac{dp_{\mathsf{m}^{\ast }}(t)}{dt}=\rho \left\{ (%
\mathsf{m}^{\ast }+1)p_{\mathsf{m}^{\ast }\mathsf{+1}}(0)+\exp \left[ -\beta %
\right] \mathsf{m}^{\ast }p_{\mathsf{m}^{\ast }\mathsf{-1}}(0)\right\} \geq 0
\end{equation*}%
for $\mathsf{m}^{\ast }\neq 0$, again contradicting $p_{\mathsf{m}^{\ast
}}(t^{\ast })<0$. Consequently we have $p_{\mathsf{m}}(t)\geq 0$ for every $%
\mathsf{m}$ and all sufficiently small $t\geq 0$.

It remains to prove the second relation in (\ref{probabilitybis}). We first
note that the inequality%
\begin{equation}
\left\vert p_{\mathsf{m}}\left( t\right) -p_{\beta ,\mathsf{Gibbs,m}%
}\right\vert \leq \exp \left[ -\frac{\beta \mathsf{m}}{2}\right] \left\Vert 
\mathsf{p}^{\ast }\right\Vert _{2,w_{\beta }}  \label{expdeaybis}
\end{equation}%
holds for each $\mathsf{m}\in \mathbb{N}$ uniformly in $t\in \left[
0,+\infty \right) $ as a consequence of (\ref{expdecay}). Then, taking (\ref%
{gibbs}) into account we obtain%
\begin{equation*}
\left\vert p_{\mathsf{m}}\left( t\right) \right\vert \leq \exp \left[ -\frac{%
\beta \mathsf{m}}{2}\right] \left\Vert \mathsf{p}^{\ast }\right\Vert
_{2,w_{\beta }}+Z_{\beta }^{-1}\exp \left[ -\beta \mathsf{m}\right]
\end{equation*}%
uniformly in $t$ for each $\mathsf{m}$, so that the series%
\begin{equation}
\sum_{\mathsf{m=0}}^{+\infty }p_{\mathsf{m}}\left( t\right) \leq Z_{\frac{%
\beta }{2}}\left\Vert \mathsf{p}^{\ast }\right\Vert _{2,w_{\beta }}+1<+\infty
\label{unifconvergence1}
\end{equation}%
converges absolutely and uniformly on $\left( 0,+\infty \right) $. Now let $%
t\in $ $\mathsf{K}\subset \left( 0,+\infty \right) $ where $\mathsf{K}$
stands for any compact subset of the interval. From (\ref{otherdecay}) we
then have, with $c_{\beta ,\rho ,\mathsf{K}}>0$ depending solely on $\beta
,\rho $ and $\mathsf{K}$,%
\begin{equation*}
\left\vert \frac{dp_{\mathsf{m}}(t)}{dt}\right\vert \leq c_{\beta ,\rho
}\exp \left[ -\frac{\beta \mathsf{m}}{2}\right] t^{-1}\left\Vert \mathsf{p}%
^{\ast }\right\Vert _{2,w_{\beta }}\leq c_{\beta ,\rho ,\mathsf{K}}\exp %
\left[ -\frac{\beta \mathsf{m}}{2}\right] \left\Vert \mathsf{p}^{\ast
}\right\Vert _{2,w_{\beta }}
\end{equation*}%
uniformly in $t$ since this variable is bounded below by the positive
distance between $\mathsf{K}$ and the origin. Consequently, the series%
\begin{equation}
\sum_{\mathsf{m=0}}^{+\infty }\left\vert \frac{dp_{\mathsf{m}}(t)}{dt}%
\right\vert \leq c_{\beta ,\rho ,\mathsf{K}}Z_{\frac{\beta }{2}}\left\Vert 
\mathsf{p}^{\ast }\right\Vert _{2,w_{\beta }}<+\infty
\label{unifconvergence2}
\end{equation}%
converges uniformly on every compact subset $\mathsf{K}$ $\subset \left(
0,+\infty \right) $. Relations (\ref{unifconvergence1}) and (\ref%
{unifconvergence2}) then imply that the function $t\rightarrow \sum_{\mathsf{%
m=0}}^{+\infty }p_{\mathsf{m}}\left( t\right) $ is differentiable on $\left(
0,+\infty \right) $, and that%
\begin{equation*}
\frac{d}{dt}\sum_{\mathsf{m=0}}^{+\infty }p_{\mathsf{m}}\left( t\right)
=\sum_{\mathsf{m=0}}^{+\infty }\frac{dp_{\mathsf{m}}(t)}{dt}=\sum_{\mathsf{%
m=0}}^{+\infty }\left( R^{\ast }\mathsf{p}\left( t\right) \right) _{\mathsf{m%
}}=0,
\end{equation*}%
where the last two equalities follow directly from (\ref{masterequationster}%
). This proves the second relation in (\ref{probabilitybis}) since $\sum_{%
\mathsf{m=0}}^{+\infty }p_{_{\mathsf{m}}}\left( 0\right) =1$ by hypothesis.
\ \ $\blacksquare $

\bigskip

\textsc{Remarks.} (1) To sum up, the oscillator initially steered away from
thermodynamical equilibrium due to its interaction with the bath will
eventually return there exponentially rapidly, as its initial probability
distribution evolves according to (\ref{masterequationster}) as a
probability distribution for all times converging toward the Gibbs
equilibrium distribution. Once thermodynamical equilibrium sets in, and
owing to the physical significance of the parameter $\rho $ in the above
equations, it is possible to retrieve several known physical results, among
which Planck's black body radiation law in case the radiation field is a
quantized electromagnetic field (see, e.g., Chapter VI of \cite{vankampen}).

(2) Whereas the semigroup structure set forth in the preceding
considerations is a manifestation of the Markovian nature of the evolution,
it is well known that the approach to equilibrium in statistical mechanics
is not Markovian in general, as was stressed in the introduction. We refer
the reader for example to \cite{emch} and \cite{radin} for very nice
examples of this fact.

\bigskip

In the next section we establish an elementary connection between the
spectral properties of the operator $R^{\ast }$ and those of a related
linear pencil (for an introduction to the spectral theory of general pencils
we refer the reader to \cite{markus}).

\section{On the spectral properties of a linear pencil associated with $%
R^{\ast }$}

We begin with the following:

\bigskip

\textbf{Proposition 6.} \textit{There exist linear bounded operators }$%
U,V:l_{\mathbb{C},w_{\beta }\text{ }}^{2}\rightarrow h_{\mathbb{C},w_{\beta
}}^{1}$\textit{\ such that}%
\begin{equation}
\left( R\mathsf{p},\mathsf{q}\right) _{2,w_{\beta }}=\left( \mathsf{p},U%
\mathsf{q}\right) _{1,2,w_{\beta }}  \label{equality}
\end{equation}%
\textit{and}%
\begin{equation}
\left( \mathsf{p},\mathsf{q}\right) _{2,w_{\beta }}=\left( \mathsf{p},V%
\mathsf{q}\right) _{1,2,w_{\beta }}  \label{equalityquarto}
\end{equation}%
\textit{for each }$\mathsf{p}\in D(R)$ \textit{and every} $\mathsf{q}\in l_{%
\mathbb{C},w_{\beta }}^{2}$\textit{, respectively, and }$V^{-1}$ \textit{%
exists as an unbounded operator from }$\func{Ran}V$ \textit{onto} $l_{%
\mathbb{C},w_{\beta }}^{2}$. \textit{Furthermore, the restrictions }$\tilde{U%
},\tilde{V}$\textit{\ of }$U,V$\textit{\ to }$h_{\mathbb{C},w_{\beta }}^{1}$%
\textit{\ are compact and self-adjoint. Finally, we have the representation}%
\begin{equation}
R^{\ast }=V^{-1}\tilde{U}  \label{extension}
\end{equation}%
\textit{of the operator} $R^{\ast }$\textit{.}

\bigskip

\textbf{Proof. }We could argue by means of the Riesz-Fr\'{e}chet
representation theorem to prove (\ref{equality}) and (\ref{equalityquarto}),
but we prefer an alternative way in order to get explicit expressions for
the operators involved. On the one hand, it is sufficient to prove those
equalities for $\mathsf{p=f}_{\mathsf{m}}$ and every $\mathsf{m}$, that is, 
\begin{equation}
\left( R\mathsf{f}_{\mathsf{m}},\mathsf{q}\right) _{2,w_{\beta }}=\left( 
\mathsf{f}_{\mathsf{m}},U\mathsf{q}\right) _{1,2,w_{\beta }}
\label{equalitybis}
\end{equation}%
and%
\begin{equation*}
\left( \mathsf{f}_{\mathsf{m}},\mathsf{q}\right) _{2,w_{\beta }}=\left( 
\mathsf{f}_{\mathsf{m}},V\mathsf{q}\right) _{1,2,w_{\beta }},
\end{equation*}%
respectively. From (\ref{sesquilinearform}) and the definition of the
weights we first have%
\begin{equation}
\left( \mathsf{f}_{\mathsf{m}},\mathsf{u}\right) _{1,2,w_{\beta }}=\exp %
\left[ \frac{\beta }{2}\mathsf{m}\right] \left( 1+\mathsf{m}^{2}\right) \bar{%
u}_{\mathsf{m}}  \label{innerproductter}
\end{equation}%
for any $\mathsf{u}\in h_{\mathbb{C},w_{\beta }}^{1}$. On the other hand,
from (\ref{innerproductbis}) we have%
\begin{eqnarray*}
&&\left( R\mathsf{f}_{\mathsf{m}},\mathsf{q}\right) _{2,w_{\beta }} \\
&=&\rho \exp \left[ \frac{\beta }{2}\mathsf{m}\right] \left\{ \left( \mathsf{%
m}+1\right) \bar{q}_{\mathsf{m}+1}+\exp \left[ -\beta \right] \mathsf{m}\bar{%
q}_{\mathsf{m}-1}-\left( \mathsf{m}+\exp \left[ -\beta \right] \left( 
\mathsf{m}+1\right) \right) \bar{q}_{\mathsf{m}}\right\}
\end{eqnarray*}%
for every $\mathsf{q}\in l_{\mathbb{C},w_{\beta }}^{2}.$ Requiring equality
between the preceding expression and (\ref{innerproductter}) then allows us
to solve for $u_{\mathsf{m}}$ and get%
\begin{equation}
u_{\mathsf{m}}=\rho \left( \frac{\mathsf{m}+1}{\mathsf{m}^{2}+1}q_{\mathsf{m}%
+1}+\frac{\exp \left[ -\beta \right] \mathsf{m}}{\mathsf{m}^{2}+1}q_{\mathsf{%
m}-1}-\frac{\mathsf{m}+\exp \left[ -\beta \right] \left( \mathsf{m}+1\right) 
}{\mathsf{m}^{2}+1}q_{\mathsf{m}}\right)  \label{equalityter}
\end{equation}%
for each $\mathsf{m}$, from which we have%
\begin{eqnarray*}
&&\left( 1+\mathsf{m}^{2}\right) \left\vert u_{\mathsf{m}}\right\vert ^{2} \\
&\leq &c_{\rho }\left( \frac{\left( \mathsf{m}+1\right) ^{2}}{\mathsf{m}%
^{2}+1}\left\vert q_{\mathsf{m}+1}\right\vert ^{2}+\frac{\exp \left[ -2\beta %
\right] \mathsf{m}^{2}}{\mathsf{m}^{2}+1}\left\vert q_{\mathsf{m}%
-1}\right\vert ^{2}+\frac{\left( \mathsf{m}+\exp \left[ -\beta \right]
\left( \mathsf{m}+1\right) \right) ^{2}}{\mathsf{m}^{2}+1}\left\vert q_{%
\mathsf{m}}\right\vert ^{2}\right)
\end{eqnarray*}%
for some irrelevant constant $c_{\rho }>0$, where the coefficients of each
term on the right-hand side are uniformly bounded in $\mathsf{m}$. We
therefore obtain%
\begin{equation*}
\left\Vert \mathsf{u}\right\Vert _{1,2,w_{\beta }}\leq c_{\beta ,\rho
}\left\Vert \mathsf{q}\right\Vert _{2,w_{\beta }}
\end{equation*}%
for some $c_{\beta ,\rho }>0$, so that 
\begin{equation*}
U:\mathsf{q}\rightarrow U\mathsf{q:}=\mathsf{u}
\end{equation*}%
is indeed a linear bounded operator from $l_{\mathbb{C},w_{\beta }\text{ }%
}^{2}$into $h_{\mathbb{C},w_{\beta }}^{1}$. The same conclusion holds for%
\begin{equation*}
V:\mathsf{q}\rightarrow V\mathsf{q:}=\mathsf{v}
\end{equation*}%
where%
\begin{equation}
v_{\mathsf{m}}=\frac{q_{\mathsf{m}}}{\mathsf{m}^{2}+1},
\label{equalityquinto}
\end{equation}%
and (\ref{equalityquinto}) implies $\ker V=\left\{ 0\right\} $ so that $%
V^{-1}$ exists as an unbounded operator from $\func{Ran}V$ onto $l_{\mathbb{C%
},w_{\beta }\text{ }}^{2}$. Furthermore, the restrictions $\tilde{U},\tilde{V%
}$ of $U,V$ to $h_{\mathbb{C},w_{\beta }}^{1}$ are compact as a consequence
of (\ref{embeddingbis}) and we omit the elementary proofs of their
self-adjointness. Finally, it is easily verified from (\ref{adjointpremier}%
), (\ref{equalityter}) and (\ref{equalityquinto}) that if $\mathsf{q}\in h_{%
\mathbb{C},w_{\beta }}^{1}$ then $U\mathsf{q=}\tilde{U}\mathsf{q}\in \func{%
Ran}V$ with%
\begin{equation}
\left( R^{\ast }\mathsf{q}\right) _{\mathsf{m}}=\left( V^{-1}\tilde{U}%
\mathsf{q}\right) _{\mathsf{m}}  \label{representation}
\end{equation}%
for every $\mathsf{m}\in \mathbb{N}$, hence $R^{\ast }=V^{-1}\tilde{U}$ on $%
h_{\mathbb{C},w_{\beta }}^{1}$. \ \ $\blacksquare $

\bigskip

Let us now consider the linear pencil%
\begin{equation*}
L\left( \nu \right) :=\tilde{U}\mathsf{-}\nu \tilde{V}
\end{equation*}%
on $h_{\mathbb{C},w_{\beta }}^{1}$, where $\nu \in \mathbb{C}$. Following
Section 11.2 in \cite{markus} or Section 1 in \cite{tretter}, we define the
spectrum of $L$, $\mathsf{\sigma (}L\mathsf{)}$, as the set of all $\nu \in 
\mathbb{C}$ for which $L\left( \nu \right) $ is not bijective, its resolvent
set as $\mathbb{C}\left\backslash \mathsf{\sigma (}L\mathsf{)}\right. $ and
its point spectrum, $\mathsf{\sigma }_{\mathsf{p}}\mathsf{(}L\mathsf{)}$, as
the set of all $\nu \in \mathbb{C}$ for which $L\left( \nu \right) $ is not
injective. Then we have:

\bigskip

\textbf{Corollary 2.} \textit{The following statements hold:}

\textit{(a) We have }$\mathsf{\sigma }(L)=$\textit{\ }$\mathbb{C}$\textit{,
that is, the resolvent set of }$L$ \textit{is empty.}

\textit{(b) The point spectrum consists exclusively of the values} \textit{%
that appear in (\ref{ordering}), that is,}%
\begin{equation*}
\mathsf{\sigma }_{\mathsf{p}}(L)=\left\{ \nu \in \mathbb{C}:\nu =\nu _{%
\mathsf{k}}\text{, }\mathsf{k}\in \mathbb{N}\text{ }\right\} ,
\end{equation*}%
\textit{which satisfy (\ref{dimension}) and correspond to the basis
eigenvectors }$\mathsf{\hat{p}}_{\mathsf{k}}$\textit{\ in (\ref%
{eigenequationbis}). Moreover we have}%
\begin{equation}
L\left( \nu _{\mathsf{k}}\right) \mathsf{\hat{p}}_{\mathsf{k}}=0
\label{eigenequationter}
\end{equation}%
\textit{for every} $\mathsf{k}\in \mathbb{N}$.

\bigskip

\textbf{Proof.} In order to prove statement (a), we first remark that for
each $\mathsf{q}\in $ $h_{\mathbb{C},w_{\beta }}^{1}$ and every $\nu \in 
\mathbb{C}$ we have $L\left( \nu \right) \mathsf{q}\in \func{Ran}V$ as a
consequence of the argument given at the very end of the proof of
Proposition 6. Furthermore, the inclusion $\func{Ran}V\subset $ $h_{\mathbb{C%
},w_{\beta }}^{1}$ is proper. For instance, the sequence $\mathsf{p}$
defined by%
\begin{equation*}
p_{\mathsf{m}}=\frac{\exp \left[ -\frac{\beta }{2}\mathsf{m}\right] }{%
\mathsf{m}^{2}+1}
\end{equation*}%
for every $\mathsf{m}$ is easily seen to belong to $h_{\mathbb{C},w_{\beta
}}^{1}$, but not to $\func{Ran}V$ according to (\ref{equalityquinto}).
Therefore, given such a $\mathsf{p}$ there exists no $\mathsf{q}\in $ $h_{%
\mathbb{C},w_{\beta }}^{1}$ satisfying $L\left( \nu \right) \mathsf{q}=%
\mathsf{p}$ so that $L\left( \nu \right) $ is not surjective.

As for Statement (b), it is plain that (\ref{eigenequationbis}) implies (\ref%
{eigenequationter}) as a consequence of the representation (\ref%
{representation}). Therefore we have the inclusion%
\begin{equation*}
\left\{ \nu \in \mathbb{C}:\nu =\nu _{\mathsf{k}}\text{, }\mathsf{k}\in 
\mathbb{N}\text{ }\right\} \subseteq \mathsf{\sigma }_{\mathsf{p}}(L).
\end{equation*}%
Now let $\hat{\nu}\in \mathsf{\sigma }_{\mathsf{p}}(L)$, so that there
exists a $\mathsf{\hat{p}}\in $ $h_{\mathbb{C},w_{\beta }}^{1}$, $\mathsf{%
\hat{p}\neq 0}$, satisfying%
\begin{equation*}
L\left( \hat{\nu}\right) \mathsf{\hat{p}}=\tilde{U}\mathsf{\hat{p}-}\hat{\nu}%
\tilde{V}\mathsf{\hat{p}=0.}
\end{equation*}%
Multiplying the preceding expression on the left by $V^{-1}$ gives%
\begin{equation*}
\left( R^{\ast }-\hat{\nu}\right) \mathsf{\hat{p}=0}
\end{equation*}%
according to (\ref{representation}), hence the converse inclusion%
\begin{equation*}
\mathsf{\sigma }_{\mathsf{p}}(L)\subseteq \left\{ \nu \in \mathbb{C}:\nu
=\nu _{\mathsf{k}}\text{, }\mathsf{k}\in \mathbb{N}\text{ }\right\}
\end{equation*}%
also holds by virtue of the spectral properties of $R^{\ast }$ proved in the
preceding section.\ \ $\blacksquare $

\bigskip

We conclude this article with two appendices:

\bigskip

\textbf{Appendix A. On the compactness of the embedding }$h_{\mathbb{C}%
,w_{\beta }}^{1}\hookrightarrow l_{\mathbb{C},w_{\beta }}^{2}$

In this appendix we derive (\ref{embeddingter}) as the consequence of a more
general embedding result that involves an intermediary space between $h_{%
\mathbb{C},w_{\beta }}^{1}$and $l_{\mathbb{C},w_{\beta }}^{2}$, and also
prove the compactness of the set $D_{\theta }$ defined in (\ref{set}). Let $%
l_{\mathbb{C},w_{\frac{\beta }{2}}}^{1}$ be the Banach space consisting of
all complex sequences $\mathsf{p=(}p_{\mathsf{m}}\mathsf{)}$ satisfying%
\begin{equation}
\left\Vert \mathsf{p}\right\Vert _{1,w_{\frac{\beta }{2}}}:=\sum_{\mathsf{m}%
=0}^{+\infty }w_{\frac{\beta }{2},\mathsf{m}}\left\vert p_{\mathsf{m}%
}\right\vert <+\infty .  \label{normbis}
\end{equation}%
Then the following result holds:

\bigskip

\textbf{Proposition A.1.} \textit{We have the embeddings}%
\begin{equation}
h_{\mathbb{C},w_{\beta }}^{1}\hookrightarrow l_{\mathbb{C},w_{\frac{\beta }{2%
}}}^{1}\rightarrow l_{\mathbb{C},w_{\beta }}^{2}  \label{embedding}
\end{equation}%
\textit{where the first one is compact and the second one continuous. In
particular, the embedding}%
\begin{equation}
h_{\mathbb{C},w_{\beta }}^{1}\hookrightarrow l_{\mathbb{C},w_{\beta }}^{2}
\label{embeddingbis}
\end{equation}%
\textit{is compact.}

\bigskip \textbf{Proof. }Remembering that $w_{\beta ,\mathsf{m}}=\exp \left[
\beta \mathsf{m}\right] $ for every $\mathsf{m}$ we first have%
\begin{eqnarray}
&&\sum_{\mathsf{m}=1}^{+\infty }w_{\frac{\beta }{2},\mathsf{m}}\left\vert p_{%
\mathsf{m}}\right\vert  \notag \\
&=&\sum_{\mathsf{m}=1}^{+\infty }\frac{1}{\mathsf{m}}w_{\frac{\beta }{2},%
\mathsf{m}}\mathsf{m}\left\vert p_{\mathsf{m}}\right\vert \leq \left( \sum_{%
\mathsf{m}=1}^{+\infty }\frac{1}{\mathsf{m}^{2}}\right) ^{\frac{1}{2}}\left(
\sum_{\mathsf{m}=1}^{+\infty }w_{\beta ,\mathsf{m}}\mathsf{m}^{2}\left\vert
p_{\mathsf{m}}\right\vert ^{2}\right) ^{\frac{1}{2}}  \label{estimatequinto}
\\
&=&\frac{\pi \sqrt{6}}{6}\left( \sum_{\mathsf{m}=1}^{+\infty }w_{\beta ,%
\mathsf{m}}\mathsf{m}^{2}\left\vert p_{\mathsf{m}}\right\vert ^{2}\right) ^{%
\frac{1}{2}}\leq \frac{\pi \sqrt{6}}{6}\left( \sum_{\mathsf{m}=0}^{+\infty
}w_{\beta ,\mathsf{m}}\left( 1+\mathsf{m}^{2}\right) \left\vert p_{\mathsf{m}%
}\right\vert ^{2}\right) ^{\frac{1}{2}}  \notag
\end{eqnarray}%
by Schwarz inequality and summing the first series on the right-hand side.
Furthermore we have%
\begin{equation*}
\left\vert p_{\mathsf{0}}\right\vert \leq \left( \sum_{\mathsf{m}%
=0}^{+\infty }w_{\beta ,\mathsf{m}}\left( 1+\mathsf{m}^{2}\right) \left\vert
p_{\mathsf{m}}\right\vert ^{2}\right) ^{\frac{1}{2}}
\end{equation*}%
and consequently%
\begin{equation*}
\left\Vert \mathsf{p}\right\Vert _{1,w_{\frac{\beta }{2}}}\leq \left( 1+%
\frac{\pi \sqrt{6}}{6}\right) \left\Vert \mathsf{p}\right\Vert
_{1,2,w_{\beta }}
\end{equation*}%
from (\ref{normter}), (\ref{normbis}) and (\ref{estimatequinto}), so that
the first embedding in (\ref{embedding}) is continuous.

Now let $\mathcal{K}$ be a bounded set in $h_{\mathbb{C},w_{\beta }}^{1}$, $%
\mathcal{\bar{K}}$ its closure in $l_{\mathbb{C},w_{\frac{\beta }{2}}}^{1}$
and let $\varepsilon >0$. Moreover, let $\kappa >0$ be the radius of a ball
centered at zero containing $\mathcal{K}$ in $h_{\mathbb{C},w_{\beta }}^{1}$%
. In order to prove that $\mathcal{\bar{K}}$ is compact in $l_{\mathbb{C},w_{%
\frac{\beta }{2}}}^{1}$, it is then necessary and sufficient to show that
there exists a finite-dimensional subspace $\mathcal{S}_{\varepsilon ,\kappa
}\subset $ $l_{\mathbb{C},w_{\frac{\beta }{2}}}^{1}$ such that the distance
of every $\mathsf{p\in }$ $\mathcal{\bar{K}}$ to $\mathcal{S}_{\varepsilon
,\kappa }$ satisfies%
\begin{equation*}
\func{dist}\left( \mathsf{p,}\mathcal{S}_{\varepsilon ,\kappa }\right)
:=\inf_{\mathsf{q\in }\mathcal{S}_{\varepsilon ,\kappa }}\left\Vert \mathsf{%
p-q}\right\Vert _{1,w_{\frac{\beta }{2}}}\leq \varepsilon ,
\end{equation*}%
a property equivalent to the total boundedness of $\mathcal{\bar{K}}$ (see,
e.g., \cite{bisgard} and \cite{kirillovgvishiani}). From an estimate similar
to (\ref{estimatequinto}), we first infer that there exists an integer $%
\mathsf{N}_{\varepsilon ,\kappa }$ such that the inequality%
\begin{equation}
\sum_{\mathsf{m}=\mathsf{N}}^{+\infty }w_{\frac{\beta }{2},\mathsf{m}%
}\left\vert p_{\mathsf{m}}\right\vert \leq \kappa \left( \sum_{\mathsf{m}=%
\mathsf{N}}^{+\infty }\frac{1}{\mathsf{m}^{2}}\right) ^{\frac{1}{2}}\leq
\kappa \frac{\varepsilon }{2\kappa }=\frac{\varepsilon }{2}
\label{estimatesexto}
\end{equation}%
holds for every $\mathsf{N}\geq \mathsf{N}_{\varepsilon ,\kappa }$. Then fix
such an $\mathsf{N}$ and let%
\begin{equation*}
\mathcal{S}_{\varepsilon ,\kappa }:=\limfunc{span}\left\{ \mathsf{e}_{0},...,%
\mathsf{e}_{\mathsf{N}}\right\}
\end{equation*}%
be the finite-dimensional subspace generated by the canonical sequences
defined before the statement of Proposition 1. Let us write $\mathsf{p}%
=\dsum\limits_{\mathsf{m}=0}^{+\infty }p_{\mathsf{m}}\mathsf{e}_{\mathsf{m}}$
for any given $\mathsf{p}\in \mathcal{K}$ and let $\mathsf{p}_{\mathsf{N}%
}:=\dsum\limits_{\mathsf{m}=0}^{\mathsf{N-1}}p_{\mathsf{m}}\mathsf{e}_{%
\mathsf{m}}\in \mathcal{S}_{\varepsilon ,\kappa }$. Since%
\begin{equation*}
\func{dist}\left( \mathsf{p,}\mathcal{S}_{\varepsilon ,\kappa }\right) \leq
\left\Vert \mathsf{p-q}\right\Vert _{1,w_{\frac{\beta }{2}}}
\end{equation*}%
for every $\mathsf{q}\in \mathcal{S}_{\varepsilon ,\kappa },$ we may choose
in particular $\mathsf{q=p}_{\mathsf{N}}$ and therefore obtain%
\begin{equation}
\func{dist}\left( \mathsf{p,}\mathcal{S}_{\varepsilon ,\kappa }\right) \leq
\left\Vert \mathsf{p-p}_{\mathsf{N}}\right\Vert _{1,w_{\frac{\beta }{2}%
}}=\sum_{\mathsf{m}=\mathsf{N}}^{+\infty }w_{\frac{\beta }{2},\mathsf{m}%
}\left\vert p_{\mathsf{m}}\right\vert \leq \frac{\varepsilon }{2}
\label{estimatesetimo}
\end{equation}%
according to (\ref{estimatesexto}). Finally, for any limit point $\mathsf{p}%
\in \mathcal{\bar{K}}$ there exists a sequence $\left( \mathsf{p}_{\mathsf{n}%
}\right) \subset \mathcal{K}$ such that $\left\Vert \mathsf{p-p}_{\mathsf{n}%
}\right\Vert _{1,w_{\frac{\beta }{2}}}\leq \frac{\varepsilon }{2}$ for each $%
\mathsf{n\geq }$ $\mathsf{n}_{\varepsilon }$ and some integer $\mathsf{n}%
_{\varepsilon }$, so that fixing such an $\mathsf{n}$ we get%
\begin{equation*}
\func{dist}\left( \mathsf{p,}\mathcal{S}_{\varepsilon ,\kappa }\right) \leq 
\frac{\varepsilon }{2}+\inf_{\mathsf{q\in }\mathcal{S}_{\varepsilon ,\kappa
}}\left\Vert \mathsf{p}_{\mathsf{n}}-\mathsf{q}\right\Vert _{1,w_{\frac{%
\beta }{2}}}\leq \varepsilon
\end{equation*}%
according to (\ref{estimatesetimo}), which proves the compactness of $%
\mathcal{\bar{K}}$ and thereby that of the first embedding in (\ref%
{embedding}).

As for the continuity of the second embedding, we simply note that%
\begin{equation*}
\left\Vert \mathsf{p}\right\Vert _{2,w_{\beta }}^{2}=\sum_{\mathsf{m}%
=0}^{+\infty }w_{\beta ,\mathsf{m}}\left\vert p_{\mathsf{m}}\right\vert
^{2}\leq \left( \sum_{\mathsf{m}=0}^{+\infty }w_{\frac{\beta }{2},\mathsf{m}%
}\left\vert p_{\mathsf{m}}\right\vert \right) ^{2}=\left\Vert \mathsf{p}%
\right\Vert _{1,w_{\frac{\beta }{2}}}^{2}.
\end{equation*}%
Consequently, the compactness of (\ref{embeddingbis}) is an immediate
consequence of (\ref{embedding}). \ \ $\blacksquare $

\bigskip

As a natural consequence of the preceding result we have:

\bigskip

\textbf{Proposition A.2.} \textit{The set}%
\begin{equation}
D_{\theta }=\left\{ \mathsf{q}\in h_{\mathbb{C},w_{\beta }}^{1}:\left\Vert 
\mathsf{q}\right\Vert _{2,w_{\beta }}\leq 1,\left\Vert S_{\nu }\mathsf{q}%
\right\Vert _{2,w_{\beta }}\leq \theta \right\}  \label{setbis}
\end{equation}%
\textit{with }$\theta \in \left[ 0,+\infty \right) $\textit{\ is compact in }%
$l_{\mathbb{C},w_{\beta }}^{2}$.

\bigskip

\textbf{Proof. } Let $\left( \mathsf{q}_{\mathsf{N}}\right) $ be any
sequence in $D_{\theta }$. Then $\left( \mathsf{q}_{\mathsf{N}}\right) $ is
bounded with respect to the graph-norm%
\begin{equation}
\left\Vert \mathsf{q}\right\Vert _{\mathcal{G},w_{\beta }}:=\left(
\left\Vert \mathsf{q}\right\Vert _{2,w_{\beta }}^{2}+\left\Vert S_{\nu }%
\mathsf{q}\right\Vert _{2,w_{\beta }}^{2}\right) ^{\frac{1}{2}}
\label{graphnorm}
\end{equation}%
which we may introduce in $D(S_{\nu })=h_{\mathbb{C},w_{\beta }}^{1},$ and
which is easily seen to be equivalent to (\ref{normter}). In particular,
since%
\begin{equation}
\left( S_{\nu }\mathsf{q}\right) _{\mathsf{m}}=\mathsf{-}\rho (\mathsf{m}%
+\nu _{\rho }+\exp \left[ -\beta \right] (\mathsf{m}+1))q_{\mathsf{m}}
\label{operatorbis}
\end{equation}%
for each $\mathsf{m}$, we infer from (\ref{normter}) and (\ref{operatorbis})
that%
\begin{equation*}
\left\Vert \mathsf{q}\right\Vert _{1,2,w_{\beta }}\leq c_{\beta ,\rho
}\left\Vert \mathsf{q}\right\Vert _{\mathcal{G},w_{\beta }}
\end{equation*}%
for every $\mathsf{q}\in $ $h_{\mathbb{C},w_{\beta }}^{1}$ for some $%
c_{\beta ,\rho }>0$ depending solely on $\beta $ \ and $\rho $. Therefore,
the sequence $\left( \mathsf{q}_{\mathsf{N}}\right) $ is necessarily bounded
with respect to the norm in $h_{\mathbb{C},w_{\beta }}^{1}$, and so there
exist a subsequence $\left( \mathsf{q}_{\mathsf{N}^{\prime }}\right) \subset 
$ $\left( \mathsf{q}_{\mathsf{N}}\right) $ and a $\mathsf{q}^{\ast }\in h_{%
\mathbb{C},w_{\beta }}^{1}$ such that $\mathsf{q}_{\mathsf{N}^{\prime
}}\rightharpoonup \mathsf{q}^{\ast }$ weakly in $h_{\mathbb{C},w_{\beta
}}^{1}$ as $\mathsf{N}^{\prime }\rightarrow +\infty .$ Since (\ref%
{embeddingbis}) is compact, we may then conclude that $\mathsf{q}_{\mathsf{N}%
^{\prime }}\rightarrow \mathsf{q}^{\ast }$ strongly in $l_{\mathbb{C}%
,w_{\beta }}^{2}$. As a consequence we now claim that (\ref{setbis}) is
indeed compact in $l_{\mathbb{C},w_{\beta }}^{2}$, which amounts to showing
that $\mathsf{q}^{\ast }\in D_{\theta }$. We already have $\left\Vert 
\mathsf{q}^{\ast }\right\Vert _{2,w_{\beta }}\leq 1$ since $\left\Vert 
\mathsf{q}_{\mathsf{N}^{\prime }}\right\Vert _{2,w_{\beta }}\leq 1$ for
every $\mathsf{N}^{\prime }$, so that it remains to prove the inequality%
\begin{equation}
\left\Vert S_{\nu }\mathsf{q}^{\ast }\right\Vert _{2,w_{\beta }}\leq \theta .
\label{bound}
\end{equation}%
Since 
\begin{equation}
\left\Vert S_{\nu }\mathsf{q}_{\mathsf{N}^{\prime }}\right\Vert _{2,w_{\beta
}}\leq \theta  \label{boundbis}
\end{equation}%
for every $\mathsf{N}^{\prime }\in \mathbb{N}$, we have \textit{a fortiori}%
\begin{equation*}
\dsum\limits_{\mathsf{m=0}}^{\mathsf{M}}\exp \left[ \beta \mathsf{m}\right] (%
\mathsf{m}+\nu _{\rho }+\exp \left[ -\beta \right] (\mathsf{m}%
+1))^{2}\left\vert \left( \mathsf{q}_{\mathsf{N}^{\prime }}\right) _{\mathsf{%
m}}\right\vert ^{2}\leq \left( \frac{\theta }{\rho }\right) ^{2}
\end{equation*}%
for each $\mathsf{M}\in \mathbb{N}$ from the very definition of $S_{\nu }$.
Moreover, the strong convergence $\mathsf{q}_{\mathsf{N}^{\prime
}}\rightarrow \mathsf{q}^{\ast }$ in $l_{\mathbb{C},w_{\beta }}^{2}$ implies
that $\left( \mathsf{q}_{\mathsf{N}^{\prime }}\right) _{\mathsf{m}%
}\rightarrow q_{\mathsf{m}}^{\ast }$ in $\mathbb{C}$ for every $\mathsf{m}$.
Fixing $\mathsf{M}$ and letting $\mathsf{N}^{\prime }\rightarrow +\infty $
in the preceding inequality we then obtain%
\begin{equation*}
\dsum\limits_{\mathsf{m=0}}^{\mathsf{M}}\exp \left[ \beta \mathsf{m}\right] (%
\mathsf{m}+\nu _{\rho }+\exp \left[ -\beta \right] (\mathsf{m}%
+1))^{2}\left\vert q_{\mathsf{m}}^{\ast }\right\vert ^{2}\leq \left( \frac{%
\theta }{\rho }\right) ^{2},
\end{equation*}%
which shows that the sum on the left-hand side is uniformly bounded and
monotone increasing in $\mathsf{M}$. Letting $\mathsf{M}\rightarrow +\infty $
in this last expression then gives (\ref{bound}). \ \ $\blacksquare $

\bigskip

Since $S_{\nu }$ is self-adjoint and satisfies%
\begin{equation*}
\left( S_{\nu }\mathsf{q},\mathsf{q}\right) _{2,w_{\beta }}\leq 0
\end{equation*}%
for every $\mathsf{q}\in h_{\mathbb{C},w_{\beta }}^{1}$, we may invoke
Theorem XIII.64 in \cite{reedsimon} to conclude that $S_{\nu }$ has a
compact resolvent, hence that $S_{\nu }^{-1}=\left( S_{\nu }-0\right) ^{-1}$
is compact since $\nu =0$ belongs to the resolvent set of $S_{\nu }$ as a
consequence of (\ref{inverse}).

\bigskip

\textbf{Appendix B.} \textbf{On the relation between quantum dynamics and
Pauli master equations for the harmonic oscillator}

In this last appendix we briefly and informally sketch the way in which the
dynamics of a quantum harmonic oscillator out of equilibrium in contact with
a radiation field leads to master equations of the form (\ref%
{masterequations}) or (\ref{masterequationsbis}). In so doing we refer the
reader to the literature listed in the main part of the text, more
specifically to Sections 2, 3 and 4 in Chapter XVII of \cite{vankampen}.
Radiation fields such as quantized electrodynamic fields in a vacuum or
lattice vibrations in a crystal are typically modelled by an infinite system
of quantum harmonic oscillators. According to the laws of Quantum Mechanics,
the equation governing the time evolution of the density operator $\mathcal{R%
}$ of a single harmonic oscillator in contact with such a field is then of
the form%
\begin{equation*}
\frac{d\mathcal{R}\left( t\right) }{dt}=-i\left[ H_{O}+H_{B}+\gamma H_{I},%
\mathcal{R}\left( t\right) \right] _{-}
\end{equation*}%
in suitable units. In the preceding expression, $H_{O}$ stands for the
Hamiltonian operator of the single harmonic oscillator, $H_{B}$ for that of
the field acting as a bath while $H_{I}$ denotes the interaction Hamiltonian
between the two, $\gamma \in \mathbb{R}$ being a coupling constant. All four
operators are related by the usual commutator $\left[ .\right] _{-}$. More
specifically, 
\begin{equation}
H_{O}=a_{0}^{\ast }a_{0},  \label{hamiltonian1}
\end{equation}%
\begin{equation}
H_{B}=\sum_{\mathsf{n}\neq 0}\omega _{\mathsf{n}}a_{\mathsf{n}}^{\ast }a_{%
\mathsf{n}}  \label{hamiltonian2}
\end{equation}%
and%
\begin{equation*}
H_{I}=\sum_{\mathsf{n}\neq 0}\gamma _{\mathsf{n}}\left( a_{\mathsf{n}%
}a_{0}^{\ast }+a_{\mathsf{n}}^{\ast }a_{0}\right)
\end{equation*}%
in suitable units, where $a_{0},a_{0}^{\ast }$ and $a_{\mathsf{n}},a_{%
\mathsf{n}}^{\ast }$ are the usual annihilation and creation operators for $%
H_{O}$ and $H_{B}$, respectively, $\omega _{\mathsf{n}}$ the frequencies of
the bath oscillators and $\gamma _{\mathsf{n}}$ some additional coupling
constants. In (\ref{hamiltonian1}) and (\ref{hamiltonian2}) the ground state
energy is omitted as it is irrelevant for the time evolution, while $H_{O}$
has been renormalized in such a way that its spectrum corresponds to the
choice made in Section 2 immediately after the proof of Lemma 1, that is to
say the set of non-negative integers. Assuming that the bath is in
thermodynamical equilibrium at inverse temperature $\beta >0$ and putting
the oscillator in contact with it at the initial time $t=0$, it is then
possible by means of several simplifying hypotheses to eliminate the bath
variables. Among others, those hypotheses include the weak coupling limit
and the Markovian approximation, in which case one eventually obtains in
closed form the so-called \textit{quantum master equation for the reduced
density operator }$R_{O}$\textit{\ of the oscillator}, namely%
\begin{eqnarray}
&&\frac{d\mathcal{R}_{O}\left( t\right) }{dt}  \notag \\
&=&-i\left[ a_{0}^{\ast }a_{0},\mathcal{R}_{O}\left( t\right) \right] _{-}
\label{reducedoperator} \\
&&+\sigma \left( a_{0}^{\ast }\mathcal{R}_{O}\left( t\right) a_{0}-\frac{1}{2%
}a_{0}a_{0}^{\ast }\mathcal{R}_{O}\left( t\right) -\frac{1}{2}\mathcal{R}%
_{O}\left( t\right) a_{0}a_{0}^{\ast }\right)  \notag \\
&&+\rho \left( a_{0}\mathcal{R}_{O}\left( t\right) a_{0}^{\ast }-\frac{1}{2}%
a_{0}^{\ast }a_{0}\mathcal{R}_{O}\left( t\right) -\frac{1}{2}\mathcal{R}%
_{O}\left( t\right) a_{0}^{\ast }a_{0}\right) .  \notag
\end{eqnarray}%
In (\ref{reducedoperator}), the constants $\sigma $ and $\rho $ are such that%
\begin{equation*}
\frac{\sigma }{\rho }=\frac{1-\exp \left[ -\beta \right] }{\exp \left[ \beta %
\right] -1}=\exp \left[ -\beta \right] ,
\end{equation*}%
which turn out to be exactly those that appear in (\ref{rates}) and (\ref%
{relation}). In fact, it is now easy to derive (\ref{masterequationsbis})
from (\ref{reducedoperator}). Let $\left\vert \mathsf{m}\right\rangle $ be
the eigenstate of the harmonic oscillator correponding to its $\mathsf{m}^{%
\mathsf{th}}$ energy level, with $\mathsf{m}\in \mathbb{N}$, and let us
recall that%
\begin{eqnarray}
a_{0}\left\vert \mathsf{m}\right\rangle &=&\sqrt{\mathsf{m}}\left\vert 
\mathsf{m-1}\right\rangle ,  \notag \\
a_{0}^{\ast }\left\vert \mathsf{m}\right\rangle &=&\sqrt{\mathsf{m+1}}%
\left\vert \mathsf{m+1}\right\rangle ,  \label{equations} \\
a_{0}^{\ast }a_{0}\left\vert \mathsf{m}\right\rangle &=&\mathsf{m}\left\vert 
\mathsf{m}\right\rangle ,  \notag
\end{eqnarray}%
and that $a_{0}^{\ast }$ is the formal adjoint of $a_{0}$ (see, e.g.,
Chapter XII in \cite{messiah} for a very clear derivation of these facts).
Defining the probability of occupation of the $\mathsf{m}^{\mathsf{th}}$
level at time $t$ as 
\begin{equation*}
p_{\mathsf{m}}(t)=\left\langle \mathsf{m}\right\vert \mathcal{R}_{O}\left(
t\right) \left\vert \mathsf{m}\right\rangle ,
\end{equation*}%
which indeed satisfies $0\leq p_{\mathsf{m}}(t)\leq 1$ by virtue of the
standard properties of density operators, we then get successively from (\ref%
{reducedoperator}) and repeated applications of all three equations in (\ref%
{equations}) the relation 
\begin{eqnarray*}
&&\frac{dp_{\mathsf{m}}(t)}{dt} \\
&=&-i\left\langle \mathsf{m}\right\vert \left[ a_{0}^{\ast }a_{0},\mathcal{R}%
_{O}\left( t\right) \right] _{-}\left\vert \mathsf{m}\right\rangle \\
&&+\sigma \left( \left\langle \mathsf{m}\right\vert a_{0}^{\ast }\mathcal{R}%
_{O}\left( t\right) a_{0}\left\vert \mathsf{m}\right\rangle -\frac{1}{2}%
\left\langle \mathsf{m}\right\vert a_{0}a_{0}^{\ast }\mathcal{R}_{O}\left(
t\right) \left\vert \mathsf{m}\right\rangle -\frac{1}{2}\left\langle \mathsf{%
m}\right\vert \mathcal{R}_{O}\left( t\right) a_{0}a_{0}^{\ast }\left\vert 
\mathsf{m}\right\rangle \right) \\
&&+\rho \left( \left\langle \mathsf{m}\right\vert a_{0}\mathcal{R}_{O}\left(
t\right) a_{0}^{\ast }\left\vert \mathsf{m}\right\rangle -\frac{1}{2}%
\left\langle \mathsf{m}\right\vert a_{0}^{\ast }a_{0}\mathcal{R}_{O}\left(
t\right) \left\vert \mathsf{m}\right\rangle -\frac{1}{2}\left\langle \mathsf{%
m}\right\vert \mathcal{R}_{O}\left( t\right) a_{0}^{\ast }a_{0}\left\vert 
\mathsf{m}\right\rangle \right) \\
&=&\sigma \left( \mathsf{m}p_{\mathsf{m-1}}(t)-\left( \mathsf{m+1}\right) p_{%
\mathsf{m}}(t)\right) +\rho \left( \left( \mathsf{m+1}\right) p_{\mathsf{m+1}%
}(t)-\mathsf{m}p_{\mathsf{m}}(t)\right) .
\end{eqnarray*}%
But regrouping terms differently in the last line we see that this is
exactly (\ref{masterequationsbis}) or, equivalently, the system of master
equations (\ref{masterequations}) with the rates (\ref{rates}).

\bigskip

\textbf{Statements and declarations}

\bigskip

\textsc{Declaration of competing interest}: The author declares that he has
no known competing financial interests or personal relationships that could
have appeared to influence the work reported in this paper.

\bigskip

\textsc{Conflict of interest}: The author states that there is no conflict
of interest regarding the content of this paper.

\bigskip

\textsc{Acknowledgments:} The author would like to thank the referee for his
or her suggestions which helped improve the presentation of this article. He
also thanks the Funda\c{c}\~{a}o para a Ci\^{e}ncia e a Tecnologia of the
Portugu\^{e}s Government (FCT) for its financial support under grant
UIDB/04561/2020.

\bigskip

\bigskip

\bigskip

\end{document}